\documentclass[%
 reprint,
  superscriptaddress,
 amsmath,amssymb,amsfonts,
 aps,
 prl,
 floatfix,
 longbibliography
]{revtex4-2}

\usepackage[utf8]{inputenc}
\usepackage[pdftex]{graphicx} \graphicspath{{}}
\usepackage{float} \usepackage{color}
\usepackage[pdftex,colorlinks=true]{hyperref}
\usepackage{subfigure}
\hypersetup{
  colorlinks=true,
  linkcolor=blue,
  citecolor = blue,
  urlcolor=blue,
}

\usepackage{mathtools}

\DeclarePairedDelimiterX\braket[2]{\langle}{\rangle}{#1 \delimsize\vert #2}
\DeclarePairedDelimiterX\expval[3]{\langle}{\rangle}{#1 \delimsize\vert #2  \delimsize\vert #3}
\DeclarePairedDelimiterX\proj[2]{\delimsize\vert#1\rangle}{\langle#2\delimsize\vert}{ }
\usepackage{siunitx}

\usepackage{tikz}
\usepackage[T1]{fontenc}
\newcommand*\emptycirc[1][0.4ex]{\tikz\draw (0,0) circle (#1);} 
\newcommand*\fullcirc[1][0.4ex]{\tikz\fill (0,0) circle (#1);}

\begin{document}

\title{Topological quantum floating phase of dipolar bosons in an optical ladder}

\author{Henning Korbmacher}
\email{henning.korbmacher@itp.uni-hannover.de}
\affiliation{Institut f\"ur Theoretische Physik, Leibniz Universit\"at Hannover, Germany}

\author{Gustavo A. Dom\'inguez-Castro}
\affiliation{Institut f\"ur Theoretische Physik, Leibniz Universit\"at Hannover, Germany}

\author{Mateusz \L\k{a}cki} 
\affiliation{Institute of Theoretical Physics, Jagiellonian University in Krakow, ul. Lojasiewicza 11,
	30-348 Krak\'ow, Poland}
 
\author{Jakub Zakrzewski}
\email{jakub.zakrzewski@uj.edu.pl}
\affiliation{Institute of Theoretical Physics, Jagiellonian University in Krakow, ul. Lojasiewicza 11,
	30-348 Krak\'ow, Poland}
\affiliation{Mark Kac Complex Systems Research Center,  Jagiellonian University in Krakow, \L{}ojasiewicza 11, 30-348 Krak\'ow, Poland}

\author{Luis Santos}
\email{santos@itp.uni-hannover.de}
\affiliation{Institut f\"ur Theoretische Physik, Leibniz Universit\"at Hannover, Germany}


\date{\today}

\begin{abstract}
Ultracold dipolar hard-core bosons in optical ladders provide exciting possibilities for the quantum simulation of anisotropic spin ladders. We show that introducing a tilt along the rungs results in a rich phase diagram at unit rung filling. In particular, for a sufficiently strong dipolar strength, the interplay between the long-range tail of the dipolar interactions and the tilting leads to the emergence of a quantum floating phase, a critical phase with incommensurate density-density correlations. Interestingly, the floating phase is topological, constituting an intermediate gapless stage in the melting of a crystal into a gapped topological Haldane phase. This novel scenario for topological floating phases in dipolar spin ladders can be investigated in on-going experiments.
\end{abstract}

\maketitle


Ultracold gases in optical lattices offer excellent possibilities for simulating spin models relevant to quantum magnetism~\cite{Bloch2008}. For short-range interacting atoms, 
super-exchange~\cite{Trotzky2008} results in weak nearest-neighbor spin-spin interactions, typically of Heisenberg type, which for fermions lead to long-range antiferromagnetic order~\cite{Mazurenko2017}. 
Additionally, van der Waals interacting Rydberg gases naturally realize the quantum Ising model~(QIM)~\cite{Browaeys2020}. The exquisite control available in these experiments enables the exploration of various lattice topologies and dimensionalities, including optical ladders.


\begin{figure}[t!]
\centering
\includegraphics[width=0.8\columnwidth]{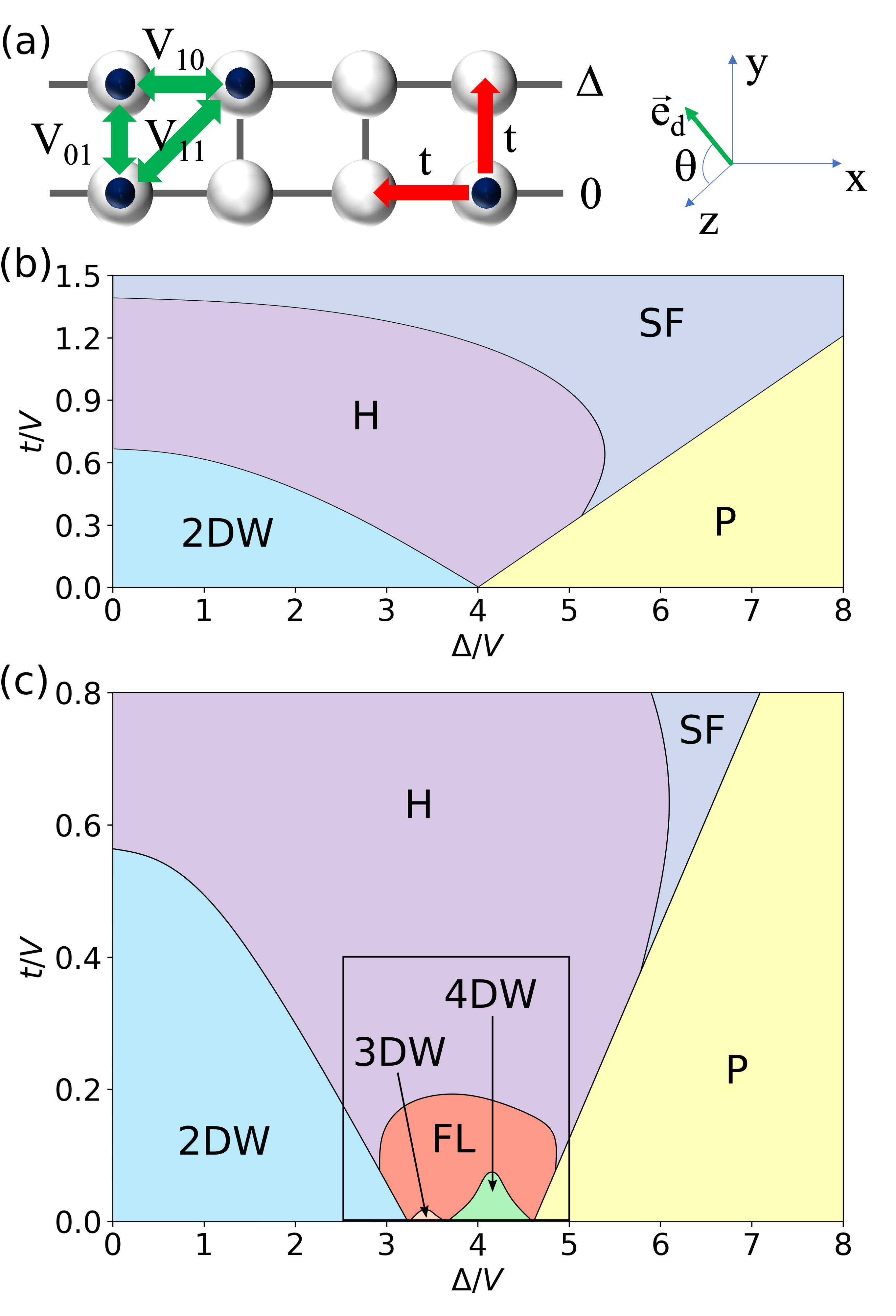}
\caption{(a) Scheme of the ladder system. (b) Phase diagram for the nearest-neighbor model showing the superfluid (SF) phase, the two-site period density wave (2DW), the polarized (P) phase, and the Haldane~(H) phase. (c) Phase diagram including the dipolar tail (up to the sixth neighbor) showing additional density modulated phases (3DW and 4DW) and the floating (FL) phase. The rectangle shows the parameter region studied in Fig.~\ref{fig:2}. The phase boundaries are evaluated using DMRG calculations with $L=96$ rungs, and monitoring the order parameters and correlations discussed in the text.}
\label{fig:1}
\end{figure}


Spin ladders play a prominent role in quantum magnetism, as an interesting intermediate case between one- and two-dimensions~\cite{Dagotto1996, Dagotto1999}. The interplay between leg and rung interactions results in rich physics for two-leg spin ladders, both in what concerns its highly non-trivial ground-state properties~\cite{Mikeska2004, Hijii2005,Li2017} 
and its dynamics~\cite{Steinigeweg2014, Schubert2021,Rakovszky2022,dominguez2023relaxation}. These physics have been the focus of recent research in optical lattice experiments, including the realization of the Haldane phase using Fermi-Hubbard ladders~\cite{Sompet2022}, or the exploration of ballistic-to-diffusive dynamics in XX spin ladders using hard-core bosons~\cite{Wienand2023}.

Recent experiments with dipolar gases, including magnetic atoms~\cite{DePaz2013, Baier2016, Fersterer2019, Su2023}, polar
molecules~\cite{Yan2013,Christakis2023,Holland2023,Bao2023,Carroll2024}, and Rydberg gases~\cite{Leseluc2019,Lienhard2020,Signoles2021,Scholl2022,Bornet2023} have realized lattice models with strong inter-site dipolar interactions, which may go well beyond nearest neighbors~\cite{Chanda2024}. These experiments open exciting possibilities for the controlled study and simulation of quantum magnetism. In particular, dipolar particles in optical ladders, formed by optical lattices or tweezers, provide a well-suited platform for the quantum simulation of fully anisotropic XXZ spin ladders.



\begin{figure*}[t!]
\centering
\includegraphics[width=0.9\linewidth]{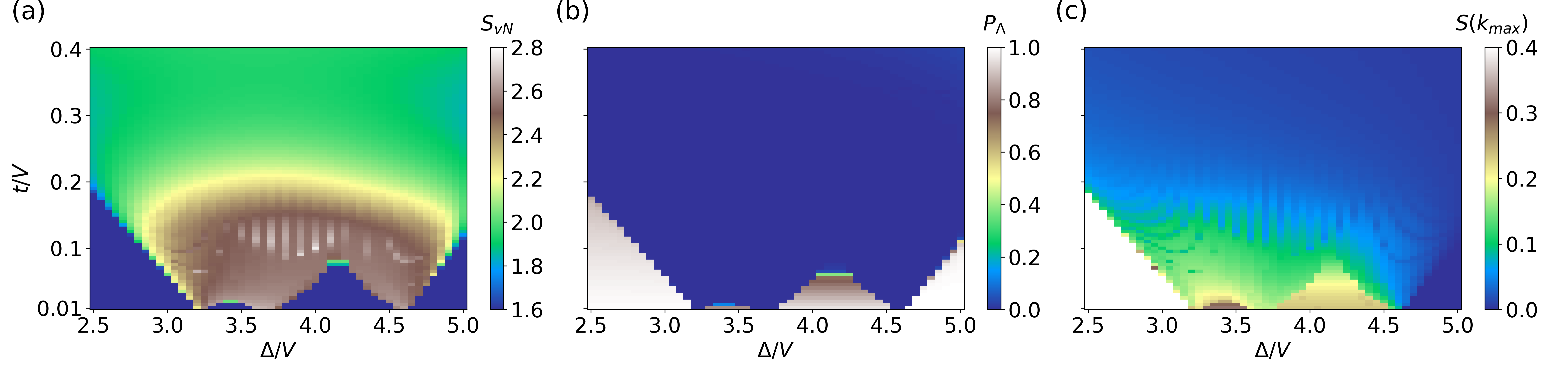}
\caption{(a) Entanglement entropy $S_{vN}$, (b) $P_\Lambda$ and (c) the maximum of the structure factor $S(k)$ 
for $k\neq 0$. Results obtained by DMRG calculations for a ladder with $L=96$ rungs.}
\label{fig:2}
\end{figure*}


This Letter illustrates some of these possibilities for the particular case of a square ladder at unit rung filling. The interplay between hopping, dipolar interactions, and a tilt of the ladder along the rung direction (readily possible in experiments) results in rich ground-state physics. Most interestingly, for finite tilting, the dipolar bosons develop a quantum  floating phase, a critical phase that is intermediate between a crystal and a disordered system with incommensurate density-density~(or equivalently spin-spin) correlations. The floating phase, originally introduced in the context of classical commensurate-incommensurate transitions in solids~\cite{Pokrovsky1979, Bak1982}, has more recently been theoretically studied in various quantum systems
~\cite{Rieger1996,Fendley2004,Weimer2010,Bermudez2012,Weimer2014, Chepiga2019, Giudici2019, Rader2019,Chepiga2021, Maceira2022}. 
Recent experiments on QIM realized with van der Waals interacting Rydberg atoms have observed the emergence of a  quantum floating phase~\cite{Zhang2024}. 

Our work shows that dipoles in ladders constitute a novel scenario in ultracold gases for the study of quantum floating phases in an anisotropic XXZ ladder. In  contrast to the QIM, the floating phase in the XXZ ladder constitutes an intermediate stage in the melting of a crystalline structure into a disordered yet symmetry-protected gapped topological Haldane phase. Interestingly, the resulting floating phase, constitutes an intriguing example of a gapless yet topological phase~\cite{Kestner2011,Ruhman2012,Thorngren2021, Fraxanet2022}.


\paragraph{Model.--}
We consider dipolar bosons in a two-leg ladder with $L$ rungs on the $xy$ plane with legs oriented along $x$~(see Fig.~\ref{fig:1}~(a)). As in recent experiments on magnetic atoms~\cite{Su2023}, we assume strongly repulsive on-site interactions that result in the hard-core condition of maximally one particle per site. The dipole moments are oriented along a direction $\vec e_d$ on the $yz$ plane forming an angle $\theta$ with the $z$-axis. For sufficiently deep lattices, the system is described by the extended Bose-Hubbard Hamiltonian
\begin{eqnarray}
    H&=&-t\sum_j\! \left ( \sum_\alpha \hat{b}^\dagger_{j,\alpha}\hat{b}_{j+1,\alpha}+\hat{b}^\dagger_{j,1}\hat{b}_{j,2}+{\mathrm H.c.}\!\right ) \! + \! \Delta\sum_j\hat{n}_{j,2}
    \nonumber\\
    &+&
    \sum_{j,\alpha}\sum_{r>0} V_{r,0}  \hat{n}_{j,\alpha}\hat{n}_{j+r,\alpha}
    + \! \sum_{j,\alpha\neq\beta}\sum_{r\geq 0}V_{r,1}  \hat{n}_{j,\alpha}\hat{n}_{j+r,\beta},
\label{eq:H}
\end{eqnarray}
where $j\in\{1,\dots,L\}$ is the rung index, $\alpha,\beta\in\{1,2\}$ are the two legs, $\hat{b}_{j,\alpha}$ ($\hat{b}^\dagger_{j,\alpha}$) is the bosonic annihilation~(creation) operator at site $j$ of leg $\alpha$, and $\hat{n}_{j,\alpha}=\hat{b}^\dagger_{j,\alpha}\hat{b}_{j,\alpha}$ is the corresponding particle number operator. For simplicity, we assume that the hopping amplitude $t$ is the same along the legs and the rungs~(although they may be in principle different~\cite{SM}). We consider leg $2$ to have a different chemical potential $\Delta>0$ compared to leg $1$. This energy bias, which may be readily realized by employing tilted ladders, plays a crucial role in the discussion below. The inter-site dipolar couplings are of the form:
\begin{equation}
V_{i_x,i_y}=
\frac{V}{\left (i_x^2 +i_y^2\right)^{3/2}}
\left(1-3\frac{ i_y^2}{i_x^2+i_y^2}\sin^2\theta\right),
\end{equation}
where $V\equiv V_{1,0}=\frac{C_{dd}}{4\pi \lambda^3}$ is the nearest-neighbor~(NN) interaction along the leg, with $\lambda$ the lattice spacing, $C_{dd}=\mu_0\mu$ for magnetic dipoles, with $\mu_0$ the vacuum permeability and $\mu$ the magnetic dipole, and $C_{dd}=d^2/\varepsilon_0$ for electric dipoles,  with $\varepsilon_0$ the vacuum dielectric constant, and $d$ the electric dipole moment. 
We are interested below in the ground-state phases at unit rung filling, i.e. when 
$\sum_{j,\alpha} \langle \hat{n}_{j,\alpha} \rangle = L$.
Due to the hard-core constraint, Eq.~\eqref{eq:H} is equivalent to the spin-$1/2$ XXZ ladder:
\begin{eqnarray}
H &=& t\sum_{j}\left [\sum_\alpha \hat{S}_{j,\alpha}^+ \hat{S}_{j+1,\alpha}^- -\hat{S}_{j,1}^+ \hat{S}_{j,2}^- + \mathrm{H. c.} \right ] - \Delta \sum_j \hat{S}_{j,2}^z
\nonumber \\
&+& \sum_{j,\alpha}\sum_{r>0} V_{r,0} \hat{S}_{j,\alpha}^z \hat{S}_{j+1,\alpha}^z
+\sum_{j,\alpha\neq \beta}\sum_{r\geq 0} V_{r,1} \hat{S}_{j,\alpha}^z \hat{S}_{j+r,\beta}^z,
\label{eq:XXZ}
\end{eqnarray}
where we introduce the spin-$1/2$ operators ${\hat S}_{j,\alpha}^+=(-1)^j {\hat b}_{j,\alpha}$ and ${\hat S}_{j\alpha}^z = 1/2-n_{j,\alpha}$, and the tilting $\Delta$ acts as an effective magnetic field gradient between the two legs. The results below are obtained by combining DMRG, with up to bond dimension $\chi=600$ and considering different $L$ with periodic boundary conditions, 
and iDMRG, with up to $\chi=300$ and considering a unit cell of $12$ rungs. In both cases we employ the TeNPy library~\cite{TeNPy2018}.



\begin{figure*}[htbp]
\centering
\includegraphics[width=1\linewidth]{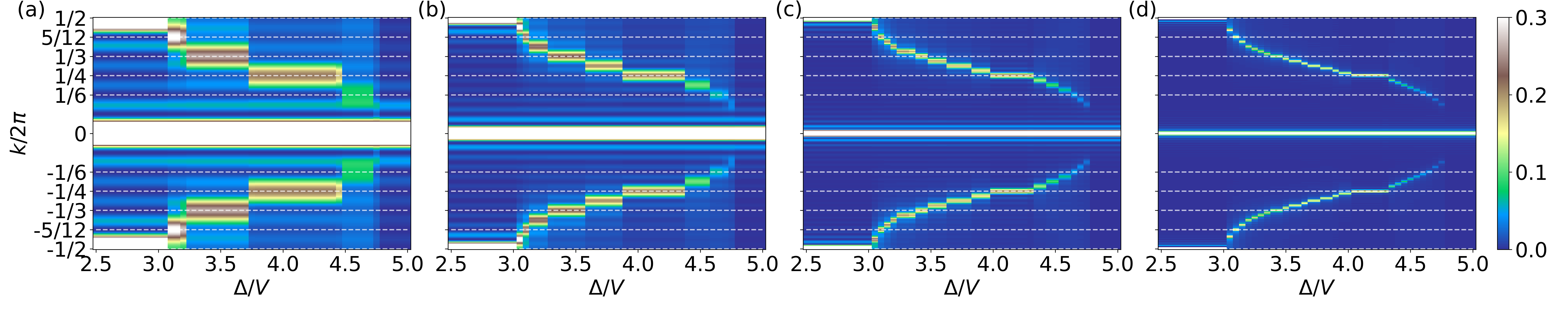}
\caption{Structure factor $S(k)$ for $t/V=0.05$ as a function of $\Delta/V$ for $L=12$~(a), $24$~(b), $48$~(c), and $96$~(d) rungs calculated using DMRG. In addition to the plateaus at $k/\pi=1$~(2DW) and $1/2$~(4DW), $S(k)$ displays peaks at intermediate incommensurate $k$ values, which 
for growing $L$ become a continuous function of $\Delta/V$, a characteristic feature of the floating phase.}
\label{fig:3}
\end{figure*}



\begin{figure}[htbp]
\centering
\includegraphics[width=0.8\columnwidth]{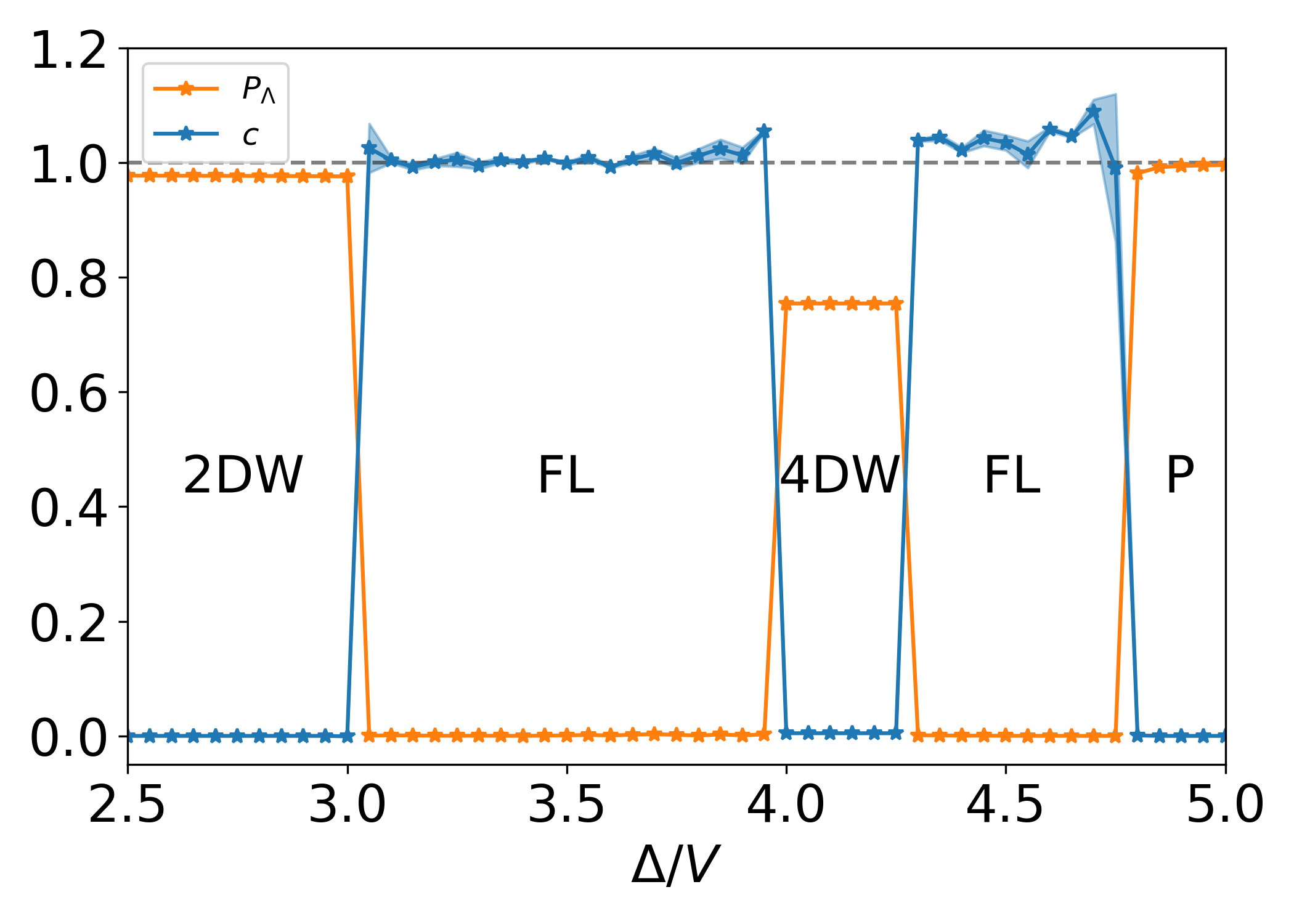}
\caption{$P_\Lambda$ (see text) and central charge $c$ obtained after a finite size scaling $\sim1/L$~(error bars indicated in light blue). Both results are obtained 
for $t/V=0.05$
from DMRG with $\chi=600$ and $L=96$ and $L\in\{60,72,84,96\}$, respectively.}
\label{fig:new}
\end{figure}




\begin{figure*}[htbp]
\centering
\includegraphics[width=0.32\linewidth]{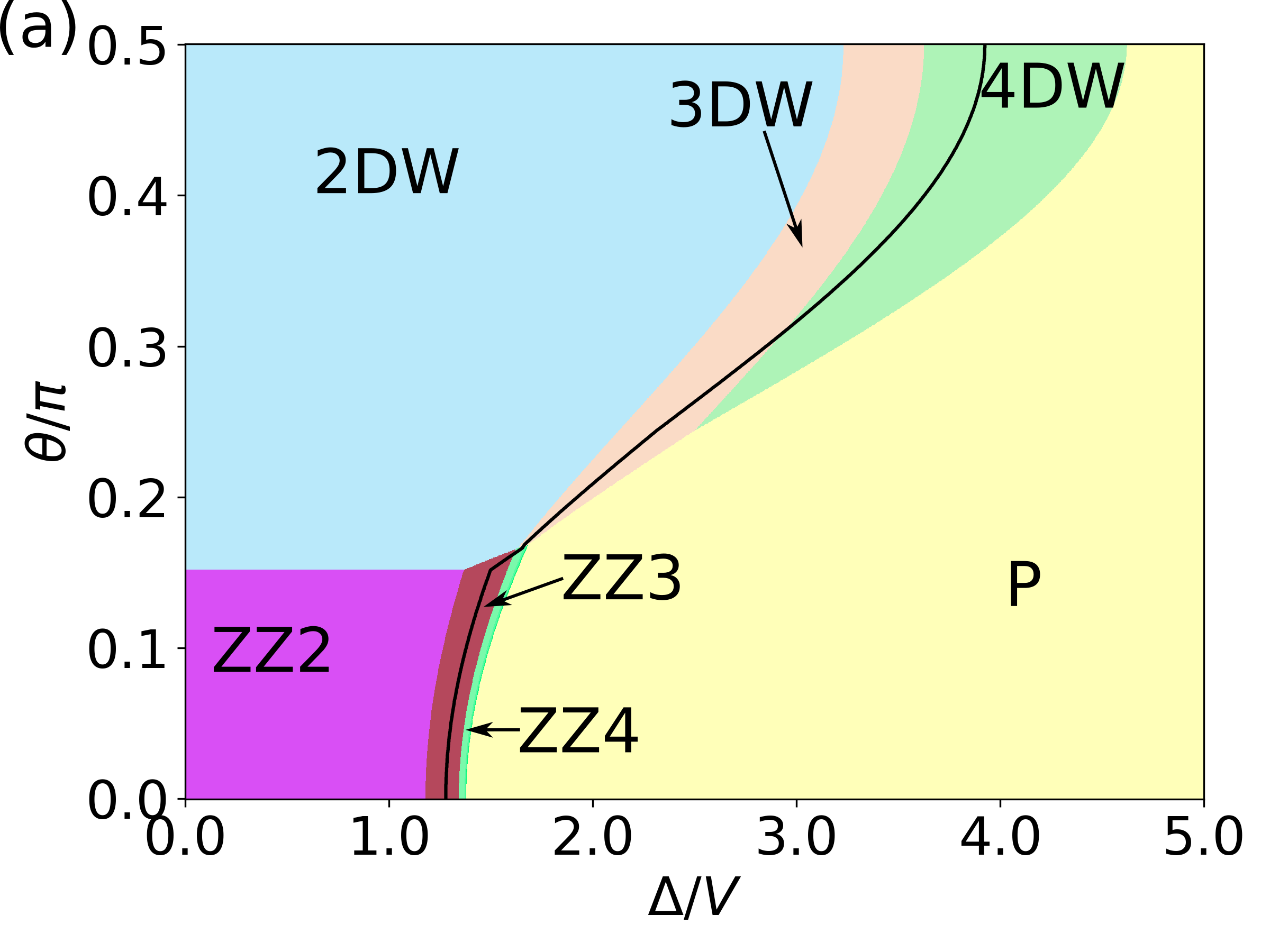}
\includegraphics[width=0.32\linewidth]{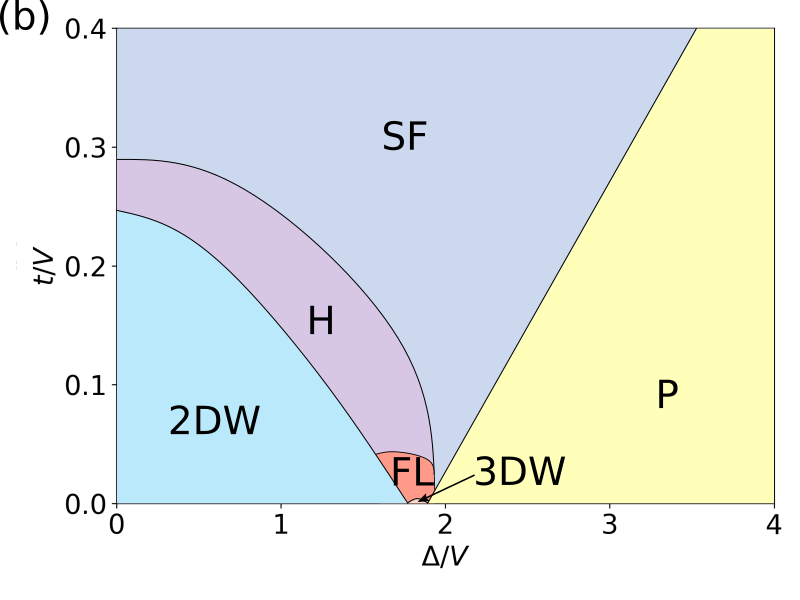}
\includegraphics[width=0.32\linewidth]{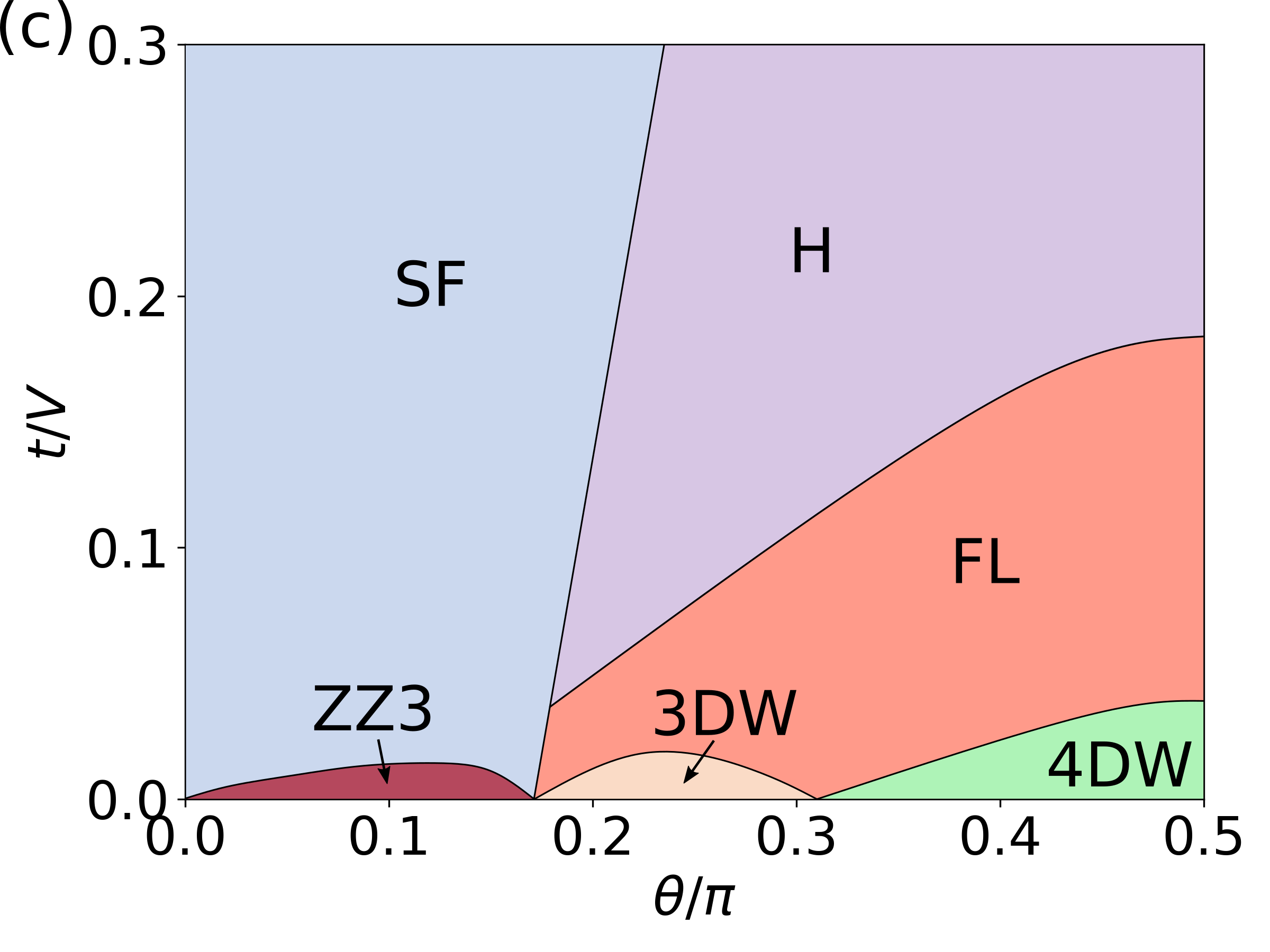}
\caption{ (a) Phases for $t=0$ as function of the dipole orientation $\theta$ and the tilting $\Delta$~(see text). The black solid curve indicates the values employed for the calculation of panel (c). (b) Phase diagram for $\theta/\pi=0.19$.
Contours drawn from iDMRG results. (c) Phase diagram along the black line panel (a).
Contours drawn from DMRG results with $L=96$ rungs.}
\label{fig:4}
\end{figure*}


\paragraph{Nearest-neighbor model.--}
We consider first dipoles oriented along $y$~($\theta=\pi/2$). We neglect at this point 
the $1/r^3$ dipolar tail, considering an NN model, with only $V_{1,0}=V$ and $V_{0,1}=-2V$. 
Due to the ferro-(antiferro-)magnetic rung~(leg) coupling, the XXZ ladder~\eqref{eq:XXZ} presents for $\Delta=0$ three phases~\cite{Hijii2005, Li2017}. Large Ising interactions~(low-enough $t/V$) result in the stripe-N\'eel phase, with staggered magnetization along the legs, i.e. a density wave of period two~(2DW) for the bosons, reflected in a peak at $k=\pi$ of the structure factor 
\begin{eqnarray}
S(k)=\frac{1}{L^2}\sum_{i,j}e^{ik(i-j)}\langle\hat{n}_i\hat{n}_j\rangle,
\end{eqnarray}
with $\hat{n}_j=\sum_\alpha \hat{n}_{j,\alpha}$. 
At $t=0$, the 2DW displays a two-rung unit cell~$\,\begin{matrix}
\vspace*{-0.18cm}
\fullcirc & \hspace*{-0.1cm}\emptycirc \\
\vspace*{0.05cm}
\fullcirc & \hspace*{-0.1cm}\emptycirc 
\end{matrix}$, with $\fullcirc$~($\emptycirc$) denoting occupied~(empty) sites.
Dominant exchange~(large-enough $t/V$) leads to the XY1 phase, a Luttinger liquid~(LL) corresponding to the superfluid~(SF) phase of the bosons, with polynomically decaying  
$\langle \hat{b}_{i,\alpha}^\dag \hat{b}_{j,\alpha} \rangle$. For intermediate $t/V$ the system is in the Haldane phase, a gapped disordered phase with non-zero string order 
\begin{equation}
\mathcal{O}_S=\lim_{|i-j|\rightarrow\infty}\left\langle \delta \hat{n}_i e^{i\pi\sum_{i<k<j}\delta \hat{n}_{k}} \delta \hat{n}_j\right\rangle, 
\end{equation} 
with $\delta \hat{n}_j=1-\hat{n}_j$. This is a symmetry protected topological phase with a doubly-degenerate entanglement spectrum~(ES)~\cite{Pollmann2010}, characterized by $P_\Lambda=0$, with 
$P_\Lambda^2\equiv \sum_{n\geq 0} |\lambda_{2n}-\lambda_{2n+1}|^2$. The ES $\{\lambda_i\}$ is given by the eigenstates of the reduced density matrix $\rho_{L/2}$
, obtained 
after cutting the ladder into two across the legs at its center, and tracing the overall density matrix over one half.

The phases are adiabatically continued for $\Delta>0$, see Fig.~\ref{fig:1}~(b). For growing $\Delta$, the 2DW phase shrinks to lower $t/V$ values until disappearing at $\Delta/V=4$. For larger $\Delta$, the system enters the polarized~(P) phase, with all particles in the lowest leg.
The Haldane phase penetrates at $\Delta/V=4$ down to $t=0$.


\paragraph{Polar lattice gas.--}  The phase diagram of the actual polar lattice gas for $\theta=\pi/2$ is shown in Fig.~\ref{fig:1}~(c). 
The  $1/r^3$ tail results in new physics in the intermediate region in between 2DW and P, see Fig.~\ref{fig:2}. 
The system displays density waves with periods $3$~(3DW) and $4$~(4DW) sites, with unit cells at $t=0$
~$\,\begin{matrix}
\vspace*{-0.18cm}
\fullcirc & \hspace*{-0.1cm}\emptycirc & \hspace*{-0.1cm}\emptycirc \\
\vspace*{0.05cm}
\fullcirc & \hspace*{-0.1cm}\fullcirc & \hspace*{-0.1cm}\emptycirc 
\end{matrix}$
and 
~$\,\begin{matrix}
\vspace*{-0.18cm}
\fullcirc & \hspace*{-0.1cm}\emptycirc & \hspace*{-0.1cm}\emptycirc & \hspace*{-0.1cm}\emptycirc \\
\vspace*{0.05cm}
\fullcirc & \hspace*{-0.1cm}\fullcirc & \hspace*{-0.1cm}\emptycirc & \hspace*{-0.1cm}\fullcirc 
\end{matrix}$, and peaks at $k=2\pi/3$ and $\pi/2$ in $S(k)$, respectively. 

\paragraph{Topological quantum floating phase.--} In stark contrast to the NN model, the Haldane phase does not penetrate down to $t=0$ through the interstitials in between the DW phases. 
For $t/V<0.2$ there is a markedly different, gapless phase, characterized by a conformal central charge $c=1$~(single component LL), see Fig.~\ref{fig:2}~(b) and Fig.~\ref{fig:new}, which we have determined both from our DMRG and iDMRG calculations from the scaling of the entanglement entropy, $S_{vN}=-\mathrm{Tr}[\rho_{L/2}\log\rho_{L/2}]$, with the system size $L$~\cite{Fehske2014} and the correlation length~\cite{Calabrese2008}, respectively. The presence of this phase 
is marked by a different entanglement entropy~(Fig.~\ref{fig:2}~(a)). More relevantly, whereas the Haldane phase is a disordered phase with no peak in $S(k\neq 0)$, the observed phase presents well-defined peaks in $S(k)$~(Fig.~\ref{fig:2}~(c)). These peaks, however, do not occur for a restricted set of discrete values linked to a well-defined density wave, but rather occur for a value of $k$ that decreases continuously from $k=\pi$ when $\Delta/V$ increases, as shown in Fig.~\ref{fig:2}~(c). This phase
hence corresponds to a quantum floating phase, for which $S(k)$ 
presents a characteristic finite-size dependence~\cite{Zhang2024}, see  Fig.~\ref{fig:3}.
The floating phase, which characterizes the melting of the DWs at small hopping, fills the whole intermediate region between 2DW and P phases, surrounding the 3DW and 4DW phases. In contrast to the recent realization in QIM ladders using Rydberg atoms~\cite{Zhang2024}, in the dipolar XXZ ladder the floating phase transitions to a disordered but topological Haldane phase. Moreover, despite of being a gapless phase, the floating phase presents both a finite string order and a double-degenerate ES, see Fig.~\ref{fig:new}. The quantum floating phase constitutes hence an intriguing example of a gapless topological phase~\cite{Kestner2011,Ruhman2012,Thorngren2021, Fraxanet2022}.


The origin of the floating phase for $t\ll V$ and $\Delta \simeq 4V$ is best understood by mapping the ladder into an effective 1D model with four possible states in a given site:
$\begin{matrix}
\vspace*{-0.18cm}
\fullcirc \\
\vspace*{0.05cm}
\fullcirc 
\end{matrix} \equiv\, 
|\!\!\uparrow\rangle$, 
$\begin{matrix}
\vspace*{-0.18cm}
\emptycirc \\
\vspace*{0.05cm}
\emptycirc 
\end{matrix} \equiv\, 
|\!\!\downarrow\rangle$,
$\begin{matrix}
\vspace*{-0.18cm}
\fullcirc \\
\vspace*{0.05cm}
\emptycirc 
\end{matrix} \equiv\, |\!\!\uparrow\downarrow\rangle$
, 
$\begin{matrix}
\vspace*{-0.18cm}
\emptycirc \\
\vspace*{0.05cm}
\fullcirc 
\end{matrix} \equiv\, |0\rangle$~(denoted as an empty site).
The 1D model is characterized by strong on-site $\uparrow$--$\downarrow$ repulsion~($2V$), and hence for $t\ll V$ 
the system reduces to a $t$-$J$-like model in which each site is either empty or with one particle in either $|\uparrow\rangle$ or $|\downarrow\rangle$. Moreover, strong $\uparrow$--$\uparrow$ and $\downarrow$--$\downarrow$ NN repulsion~($V$), results in antiferromagnetic order. Since the spin configuration is fixed, we focus on the distribution of empty sites, which we denote as holons. 
Considering for simplicity up to 
next-to-NN~(NNN) interactions, 
the holons are given by the 1D hard-core extended-Hubbard model~\cite{SM}: 
\begin{eqnarray}
    H&=&-t\sum_j \left ( \hat{c}^\dagger_{j}\hat{c}_{j+1}+\mathrm{H.c.}\!\right ) 
    - \! \mu \sum_j \hat N_j \nonumber \\
    &+& V_{NNN} \sum_j  \hat{N}_j\hat{N}_{j+1} -\frac{t^2}{V} \sum_j \left (\hat c^\dag_j \hat c^\dag _{j+1} + \mathrm{H. c.} \right ),
\label{eq:H_ph_ph}
\end{eqnarray}
where $\hat{c}^\dagger_j$ creates a holon in site $j$, $\hat{N}_j=0,1$ is the holon occupation in rung $j$, $\mu = \frac{1}{2}\left (\Delta - 4V+ 2 V_{\mathrm{NNN}} \right )$, and $V_{\mathrm{NNN}}= \frac{V}{2^{5/2}}$ is the NNN interaction. Note that we have included the second-order, but potentially relevant, term resulting from super-exchange involving a rung and a leg hopping that creates/destroys NN holons~\cite{SM}. The NNN interactions prevent the creation/destruction of NN holon pairs for $t^2/V \ll V_{\mathrm{NNN}}$. As a result, the system reduces to an extended Hubbard model 
with a well-defined holon filling, which presents four different ground states, a phase with unit filling of holons (the P phase), a phase with no holons (the 2DW), a density wave at half filling of holons~(the 4DW), and an LL with delocalized holons~(with up to NNN interactions there is no 3DW phase~\cite{SM}). Doping the 2DW with delocalized holons following a commensurate-incommensurate transition into the LL phase, results in an effectively stretched separation between particles, which translates into a peak in $S(k)$ at incommensurate $k$ values. Hence the LL of holons is the floating phase.  When $t/V$ increases, the second-order term eventually induces an efficient uncorrelated holon-pair creation/destruction, which results in exponentially decaying single-holon correlations. The system hence enters into an insulating phase, which as for the floating phase presents a diluted antiferromagnetic order, and hence finite $\cal{O}_S$.  This insulating phase is thus the Haldane insulator. 
Finally, note that for the NN model, the chemical potential $\mu\simeq 0$ when $\Delta\simeq 4V$. In the absence of NNN interactions, even a vanishing $t/V$ results in efficient uncorrelated creation/destruction of holon pairs, precluding the floating phase. Finally, note that 
observing the floating phase demands a temperature $T<\mu$ such that 
holons are not efficiently thermally created/destroyed.


\paragraph{Dependence on the dipole orientation.--} 
We considered above $\theta=\pi/2$. Figure~\ref{fig:4}~(a) shows the results for $t=0$ as a function of $\theta$ and $\Delta/V$.
For $\theta/\pi<0.17$, a value close but not equal to the magic angle $\theta_M=\arcsin(1/\sqrt{3})$ at which $V_{0,1}=0$, the 2DW transforms into a zigzag wave~(ZZ2)~\cite{Ruhman2012,Gammelmark2013}, 
that corresponds to the N\'eel phase of XXZ spin ladders, which presents also period 2, but has a  unit cell~$\,\begin{matrix}
\vspace*{-0.18cm}
\fullcirc & \hspace*{-0.1cm}\emptycirc \\
\vspace*{0.05cm}
\emptycirc & \hspace*{-0.1cm}\fullcirc 
\end{matrix}$ at $t=0$. Additional 
density waves occur, with period 3~(ZZ3) and 4~(ZZ4) with unit cell 
~$\,\begin{matrix}
\vspace*{-0.18cm}
\fullcirc & \hspace*{-0.1cm}\emptycirc & \hspace*{-0.1cm}\emptycirc \\
\vspace*{0.05cm}
\emptycirc & \hspace*{-0.1cm}\fullcirc & \hspace*{-0.1cm}\fullcirc 
\end{matrix}$
and 
$\,\begin{matrix}
\vspace*{-0.18cm}
\fullcirc & \hspace*{-0.1cm}\emptycirc & \hspace*{-0.1cm}\emptycirc & \hspace*{-0.1cm}\emptycirc \\
\vspace*{0.05cm}
\emptycirc & \hspace*{-0.1cm}\fullcirc & \hspace*{-0.1cm}\fullcirc & \hspace*{-0.1cm}\fullcirc 
\end{matrix}$, respectively. When $\theta$ decreases the Haldane phase shrinks in favor of the SF~(see Fig.~\ref{fig:3}~(b) for $\theta/\pi=0.19$). The appearance of ZZ phases 
correlates with the disappearance of the Haldane and the floating phases~(see Fig.~\ref{fig:3}~(c)).


\paragraph{Conclusions.--}
Dipolar particles in optical ladders are an interesting novel platform for the realization of a quantum floating phase, which is gapless but topological, intermediate between a crystal and a gapped topological disordered Haldane phase. Observing the floating phase demands $V/t\gtrsim 5$, although if rung hopping is weaker than leg hopping, the floating phase may appear for significantly lower $V/t$ values~\cite{SM}. 
Since on-going experiments have already achieved $V/t=6$~\cite{Su2023}, the topological floating phase in dipolar ladders is hence within experimental reach. The string-order and structure factor characterizing the floating phase may be probed by means of site-resolved quantum gas microscopy.
Our results open interesting questions concerning the phase transitions associated to crystal melting, which may present a highly non-trivial nature~\cite{Fisher1982, Fendley2004, Chepiga2019, Giudici2019, Maceira2022}, the non-equilibrium dynamics of the dipolar ladder, and multi-leg ladders.



\acknowledgments
H.K., G.A.D.-C. and L.S acknowledge the support of the Deutsche Forschungsgemeinschaft (DFG, German Research Foundation) -- Project-ID 274200144 -- SFB 1227 DQ-mat within the project A04, and under Germany's Excellence Strategy -- EXC-2123 Quantum-Frontiers -- 390837967). M.\L. acknowledges support from
the National Science Centre (Poland) via Opus Grant No. 2019/35/B/ST2/00838.
M.\L. gratefully acknowledges Polish high-performance computing infrastructure PLGrid (HPC Centers: ACK Cyfronet AGH) for providing computer facilities and support within computational grant no. PLG/2024/017242.
J.Z. acknowledges support by the National Science Centre (Poland)  under the OPUS call within the WEAVE program 2021/43/I/ST3/01142. 
The research has also been supported by the Priority Research Area (DigiWorld) under the Strategic Programme Excellence Initiative at Jagiellonian University (M.\L., J.Z.).

\bibliography{Dipolar-Ladder}

\begin{thebibliography}{54}%
\makeatletter
\providecommand \@ifxundefined [1]{%
 \@ifx{#1\undefined}
}%
\providecommand \@ifnum [1]{%
 \ifnum #1\expandafter \@firstoftwo
 \else \expandafter \@secondoftwo
 \fi
}%
\providecommand \@ifx [1]{%
 \ifx #1\expandafter \@firstoftwo
 \else \expandafter \@secondoftwo
 \fi
}%
\providecommand \natexlab [1]{#1}%
\providecommand \enquote  [1]{``#1''}%
\providecommand \bibnamefont  [1]{#1}%
\providecommand \bibfnamefont [1]{#1}%
\providecommand \citenamefont [1]{#1}%
\providecommand \href@noop [0]{\@secondoftwo}%
\providecommand \href [0]{\begingroup \@sanitize@url \@href}%
\providecommand \@href[1]{\@@startlink{#1}\@@href}%
\providecommand \@@href[1]{\endgroup#1\@@endlink}%
\providecommand \@sanitize@url [0]{\catcode `\\12\catcode `\$12\catcode
  `\&12\catcode `\#12\catcode `\^12\catcode `\_12\catcode `\%12\relax}%
\providecommand \@@startlink[1]{}%
\providecommand \@@endlink[0]{}%
\providecommand \url  [0]{\begingroup\@sanitize@url \@url }%
\providecommand \@url [1]{\endgroup\@href {#1}{\urlprefix }}%
\providecommand \urlprefix  [0]{URL }%
\providecommand \Eprint [0]{\href }%
\providecommand \doibase [0]{https://doi.org/}%
\providecommand \selectlanguage [0]{\@gobble}%
\providecommand \bibinfo  [0]{\@secondoftwo}%
\providecommand \bibfield  [0]{\@secondoftwo}%
\providecommand \translation [1]{[#1]}%
\providecommand \BibitemOpen [0]{}%
\providecommand \bibitemStop [0]{}%
\providecommand \bibitemNoStop [0]{.\EOS\space}%
\providecommand \EOS [0]{\spacefactor3000\relax}%
\providecommand \BibitemShut  [1]{\csname bibitem#1\endcsname}%
\let\auto@bib@innerbib\@empty
\bibitem [{\citenamefont {Bloch}\ \emph {et~al.}(2008)\citenamefont {Bloch},
  \citenamefont {Dalibard},\ and\ \citenamefont {Zwerger}}]{Bloch2008}%
  \BibitemOpen
  \bibfield  {author} {\bibinfo {author} {\bibfnamefont {I.}~\bibnamefont
  {Bloch}}, \bibinfo {author} {\bibfnamefont {J.}~\bibnamefont {Dalibard}},\
  and\ \bibinfo {author} {\bibfnamefont {W.}~\bibnamefont {Zwerger}},\
  }\bibfield  {title} {\bibinfo {title} {Many-body physics with ultracold
  gases},\ }\href {https://doi.org/10.1103/RevModPhys.80.885} {\bibfield
  {journal} {\bibinfo  {journal} {Rev. Mod. Phys.}\ }\textbf {\bibinfo {volume}
  {80}},\ \bibinfo {pages} {885} (\bibinfo {year} {2008})}\BibitemShut
  {NoStop}%
\bibitem [{\citenamefont {Trotzky}\ \emph {et~al.}(2008)\citenamefont
  {Trotzky}, \citenamefont {Cheinet}, \citenamefont {F\"olling}, \citenamefont
  {Feld}, \citenamefont {Schnorrberger}, \citenamefont {Rey}, \citenamefont
  {Polkovnikov}, \citenamefont {Demler}, \citenamefont {Lukin},\ and\
  \citenamefont {Bloch}}]{Trotzky2008}%
  \BibitemOpen
  \bibfield  {author} {\bibinfo {author} {\bibfnamefont {S.}~\bibnamefont
  {Trotzky}}, \bibinfo {author} {\bibfnamefont {P.}~\bibnamefont {Cheinet}},
  \bibinfo {author} {\bibfnamefont {S.}~\bibnamefont {F\"olling}}, \bibinfo
  {author} {\bibfnamefont {M.}~\bibnamefont {Feld}}, \bibinfo {author}
  {\bibfnamefont {U.}~\bibnamefont {Schnorrberger}}, \bibinfo {author}
  {\bibfnamefont {A.~M.}\ \bibnamefont {Rey}}, \bibinfo {author} {\bibfnamefont
  {A.}~\bibnamefont {Polkovnikov}}, \bibinfo {author} {\bibfnamefont {E.~A.}\
  \bibnamefont {Demler}}, \bibinfo {author} {\bibfnamefont {M.~D.}\
  \bibnamefont {Lukin}},\ and\ \bibinfo {author} {\bibfnamefont
  {I.}~\bibnamefont {Bloch}},\ }\bibfield  {title} {\bibinfo {title}
  {Time-resolved observation and control of superexchange interactions with
  ultracold atoms in optical lattices},\ }\href
  {https://doi.org/10.1126/science.1150841} {\bibfield  {journal} {\bibinfo
  {journal} {Science}\ }\textbf {\bibinfo {volume} {319}},\ \bibinfo {pages}
  {295} (\bibinfo {year} {2008})}\BibitemShut {NoStop}%
\bibitem [{\citenamefont {Mazurenko}\ \emph {et~al.}(2017)\citenamefont
  {Mazurenko}, \citenamefont {Chiu}, \citenamefont {Ji}, \citenamefont
  {Parsons}, \citenamefont {Kan{\'a}sz-Nagy}, \citenamefont {Schmidt},
  \citenamefont {Grusdt}, \citenamefont {Demler}, \citenamefont {Greif},\ and\
  \citenamefont {Greiner}}]{Mazurenko2017}%
  \BibitemOpen
  \bibfield  {author} {\bibinfo {author} {\bibfnamefont {A.}~\bibnamefont
  {Mazurenko}}, \bibinfo {author} {\bibfnamefont {C.~S.}\ \bibnamefont {Chiu}},
  \bibinfo {author} {\bibfnamefont {G.}~\bibnamefont {Ji}}, \bibinfo {author}
  {\bibfnamefont {M.~F.}\ \bibnamefont {Parsons}}, \bibinfo {author}
  {\bibfnamefont {M.}~\bibnamefont {Kan{\'a}sz-Nagy}}, \bibinfo {author}
  {\bibfnamefont {R.}~\bibnamefont {Schmidt}}, \bibinfo {author} {\bibfnamefont
  {F.}~\bibnamefont {Grusdt}}, \bibinfo {author} {\bibfnamefont
  {E.}~\bibnamefont {Demler}}, \bibinfo {author} {\bibfnamefont
  {D.}~\bibnamefont {Greif}},\ and\ \bibinfo {author} {\bibfnamefont
  {M.}~\bibnamefont {Greiner}},\ }\bibfield  {title} {\bibinfo {title} {A
  cold-atom fermi--hubbard antiferromagnet},\ }\href
  {https://doi.org/10.1038/nature22362} {\bibfield  {journal} {\bibinfo
  {journal} {Nature}\ }\textbf {\bibinfo {volume} {545}},\ \bibinfo {pages}
  {462} (\bibinfo {year} {2017})}\BibitemShut {NoStop}%
\bibitem [{\citenamefont {Browaeys}\ and\ \citenamefont
  {Lahaye}(2020)}]{Browaeys2020}%
  \BibitemOpen
  \bibfield  {author} {\bibinfo {author} {\bibfnamefont {A.}~\bibnamefont
  {Browaeys}}\ and\ \bibinfo {author} {\bibfnamefont {T.}~\bibnamefont
  {Lahaye}},\ }\bibfield  {title} {\bibinfo {title} {Many-body physics with
  individually controlled rydberg atoms},\ }\href
  {https://doi.org/10.1038/s41567-019-0733-z} {\bibfield  {journal} {\bibinfo
  {journal} {Nature Physics}\ }\textbf {\bibinfo {volume} {16}},\ \bibinfo
  {pages} {132} (\bibinfo {year} {2020})}\BibitemShut {NoStop}%
\bibitem [{\citenamefont {Dagotto}\ and\ \citenamefont
  {Rice}(1996)}]{Dagotto1996}%
  \BibitemOpen
  \bibfield  {author} {\bibinfo {author} {\bibfnamefont {E.}~\bibnamefont
  {Dagotto}}\ and\ \bibinfo {author} {\bibfnamefont {T.~M.}\ \bibnamefont
  {Rice}},\ }\bibfield  {title} {\bibinfo {title} {Surprises on the way from
  one- to two-dimensional quantum magnets: The ladder materials},\ }\href
  {https://doi.org/10.1126/science.271.5249.618} {\bibfield  {journal}
  {\bibinfo  {journal} {Science}\ }\textbf {\bibinfo {volume} {271}},\ \bibinfo
  {pages} {618} (\bibinfo {year} {1996})}\BibitemShut {NoStop}%
\bibitem [{\citenamefont {Dagotto}(1999)}]{Dagotto1999}%
  \BibitemOpen
  \bibfield  {author} {\bibinfo {author} {\bibfnamefont {E.}~\bibnamefont
  {Dagotto}},\ }\bibfield  {title} {\bibinfo {title} {Experiments on ladders
  reveal a complex interplay between a spin-gapped normal state and
  superconductivity},\ }\href {https://doi.org/10.1088/0034-4885/62/11/202}
  {\bibfield  {journal} {\bibinfo  {journal} {Reports on Progress in Physics}\
  }\textbf {\bibinfo {volume} {62}},\ \bibinfo {pages} {1525} (\bibinfo {year}
  {1999})}\BibitemShut {NoStop}%
\bibitem [{\citenamefont {Mikeska}\ and\ \citenamefont
  {Kolezhuk}(2004)}]{Mikeska2004}%
  \BibitemOpen
  \bibfield  {author} {\bibinfo {author} {\bibfnamefont {H.-J.}\ \bibnamefont
  {Mikeska}}\ and\ \bibinfo {author} {\bibfnamefont {A.~K.}\ \bibnamefont
  {Kolezhuk}},\ }\bibinfo {title} {One-dimensional magnetism},\ in\ \href
  {https://doi.org/10.1007/BFb0119591} {\emph {\bibinfo {booktitle} {Quantum
  Magnetism}}},\ \bibinfo {editor} {edited by\ \bibinfo {editor} {\bibfnamefont
  {U.}~\bibnamefont {Schollw{\"o}ck}}, \bibinfo {editor} {\bibfnamefont
  {J.}~\bibnamefont {Richter}}, \bibinfo {editor} {\bibfnamefont {D.~J.~J.}\
  \bibnamefont {Farnell}},\ and\ \bibinfo {editor} {\bibfnamefont {R.~F.}\
  \bibnamefont {Bishop}}}\ (\bibinfo  {publisher} {Springer Berlin
  Heidelberg},\ \bibinfo {address} {Berlin, Heidelberg},\ \bibinfo {year}
  {2004})\ pp.\ \bibinfo {pages} {1--83}\BibitemShut {NoStop}%
\bibitem [{\citenamefont {Hijii}\ \emph {et~al.}(2005)\citenamefont {Hijii},
  \citenamefont {Kitazawa},\ and\ \citenamefont {Nomura}}]{Hijii2005}%
  \BibitemOpen
  \bibfield  {author} {\bibinfo {author} {\bibfnamefont {K.}~\bibnamefont
  {Hijii}}, \bibinfo {author} {\bibfnamefont {A.}~\bibnamefont {Kitazawa}},\
  and\ \bibinfo {author} {\bibfnamefont {K.}~\bibnamefont {Nomura}},\
  }\bibfield  {title} {\bibinfo {title} {Phase diagram of
  $\mathrm{S}=\frac{1}{2}$ two-leg $xxz$ spin-ladder systems},\ }\href
  {https://doi.org/10.1103/PhysRevB.72.014449} {\bibfield  {journal} {\bibinfo
  {journal} {Phys. Rev. B}\ }\textbf {\bibinfo {volume} {72}},\ \bibinfo
  {pages} {014449} (\bibinfo {year} {2005})}\BibitemShut {NoStop}%
\bibitem [{\citenamefont {Li}\ \emph {et~al.}(2017)\citenamefont {Li},
  \citenamefont {Shi}, \citenamefont {Batchelor},\ and\ \citenamefont
  {Zhou}}]{Li2017}%
  \BibitemOpen
  \bibfield  {author} {\bibinfo {author} {\bibfnamefont {S.-H.}\ \bibnamefont
  {Li}}, \bibinfo {author} {\bibfnamefont {Q.-Q.}\ \bibnamefont {Shi}},
  \bibinfo {author} {\bibfnamefont {M.~T.}\ \bibnamefont {Batchelor}},\ and\
  \bibinfo {author} {\bibfnamefont {H.-Q.}\ \bibnamefont {Zhou}},\ }\bibfield
  {title} {\bibinfo {title} {Groundstate fidelity phase diagram of the fully
  anisotropic two-leg spin-1/2 xxz ladder},\ }\href
  {https://doi.org/10.1088/1367-2630/aa8bce} {\bibfield  {journal} {\bibinfo
  {journal} {New Journal of Physics}\ }\textbf {\bibinfo {volume} {19}},\
  \bibinfo {pages} {113027} (\bibinfo {year} {2017})}\BibitemShut {NoStop}%
\bibitem [{\citenamefont {Steinigeweg}\ \emph {et~al.}(2014)\citenamefont
  {Steinigeweg}, \citenamefont {Heidrich-Meisner}, \citenamefont {Gemmer},
  \citenamefont {Michielsen},\ and\ \citenamefont
  {De~Raedt}}]{Steinigeweg2014}%
  \BibitemOpen
  \bibfield  {author} {\bibinfo {author} {\bibfnamefont {R.}~\bibnamefont
  {Steinigeweg}}, \bibinfo {author} {\bibfnamefont {F.}~\bibnamefont
  {Heidrich-Meisner}}, \bibinfo {author} {\bibfnamefont {J.}~\bibnamefont
  {Gemmer}}, \bibinfo {author} {\bibfnamefont {K.}~\bibnamefont {Michielsen}},\
  and\ \bibinfo {author} {\bibfnamefont {H.}~\bibnamefont {De~Raedt}},\
  }\bibfield  {title} {\bibinfo {title} {Scaling of diffusion constants in the
  spin-$\frac{1}{2}$ xx ladder},\ }\href
  {https://doi.org/10.1103/PhysRevB.90.094417} {\bibfield  {journal} {\bibinfo
  {journal} {Phys. Rev. B}\ }\textbf {\bibinfo {volume} {90}},\ \bibinfo
  {pages} {094417} (\bibinfo {year} {2014})}\BibitemShut {NoStop}%
\bibitem [{\citenamefont {Schubert}\ \emph {et~al.}(2021)\citenamefont
  {Schubert}, \citenamefont {Richter}, \citenamefont {Jin}, \citenamefont
  {Michielsen}, \citenamefont {De~Raedt},\ and\ \citenamefont
  {Steinigeweg}}]{Schubert2021}%
  \BibitemOpen
  \bibfield  {author} {\bibinfo {author} {\bibfnamefont {D.}~\bibnamefont
  {Schubert}}, \bibinfo {author} {\bibfnamefont {J.}~\bibnamefont {Richter}},
  \bibinfo {author} {\bibfnamefont {F.}~\bibnamefont {Jin}}, \bibinfo {author}
  {\bibfnamefont {K.}~\bibnamefont {Michielsen}}, \bibinfo {author}
  {\bibfnamefont {H.}~\bibnamefont {De~Raedt}},\ and\ \bibinfo {author}
  {\bibfnamefont {R.}~\bibnamefont {Steinigeweg}},\ }\bibfield  {title}
  {\bibinfo {title} {Quantum versus classical dynamics in spin models: Chains,
  ladders, and square lattices},\ }\href
  {https://doi.org/10.1103/PhysRevB.104.054415} {\bibfield  {journal} {\bibinfo
   {journal} {Phys. Rev. B}\ }\textbf {\bibinfo {volume} {104}},\ \bibinfo
  {pages} {054415} (\bibinfo {year} {2021})}\BibitemShut {NoStop}%
\bibitem [{\citenamefont {Rakovszky}\ \emph {et~al.}(2022)\citenamefont
  {Rakovszky}, \citenamefont {von Keyserlingk},\ and\ \citenamefont
  {Pollmann}}]{Rakovszky2022}%
  \BibitemOpen
  \bibfield  {author} {\bibinfo {author} {\bibfnamefont {T.}~\bibnamefont
  {Rakovszky}}, \bibinfo {author} {\bibfnamefont {C.~W.}\ \bibnamefont {von
  Keyserlingk}},\ and\ \bibinfo {author} {\bibfnamefont {F.}~\bibnamefont
  {Pollmann}},\ }\bibfield  {title} {\bibinfo {title} {Dissipation-assisted
  operator evolution method for capturing hydrodynamic transport},\ }\href
  {https://doi.org/10.1103/PhysRevB.105.075131} {\bibfield  {journal} {\bibinfo
   {journal} {Phys. Rev. B}\ }\textbf {\bibinfo {volume} {105}},\ \bibinfo
  {pages} {075131} (\bibinfo {year} {2022})}\BibitemShut {NoStop}%
\bibitem [{\citenamefont {Dom\'{\i}nguez-Castro}\ \emph
  {et~al.}(2024)\citenamefont {Dom\'{\i}nguez-Castro}, \citenamefont
  {Bilitewski}, \citenamefont {Wellnitz}, \citenamefont {Rey},\ and\
  \citenamefont {Santos}}]{dominguez2023relaxation}%
  \BibitemOpen
  \bibfield  {author} {\bibinfo {author} {\bibfnamefont {G.~A.}\ \bibnamefont
  {Dom\'{\i}nguez-Castro}}, \bibinfo {author} {\bibfnamefont {T.}~\bibnamefont
  {Bilitewski}}, \bibinfo {author} {\bibfnamefont {D.}~\bibnamefont
  {Wellnitz}}, \bibinfo {author} {\bibfnamefont {A.~M.}\ \bibnamefont {Rey}},\
  and\ \bibinfo {author} {\bibfnamefont {L.}~\bibnamefont {Santos}},\
  }\bibfield  {title} {\bibinfo {title} {Relaxation in dipolar spin ladders:
  From pair production to false-vacuum decay},\ }\href
  {https://doi.org/10.1103/PhysRevA.110.L021302} {\bibfield  {journal}
  {\bibinfo  {journal} {Phys. Rev. A}\ }\textbf {\bibinfo {volume} {110}},\
  \bibinfo {pages} {L021302} (\bibinfo {year} {2024})}\BibitemShut {NoStop}%
\bibitem [{\citenamefont {Sompet}\ \emph {et~al.}(2022)\citenamefont {Sompet},
  \citenamefont {Hirthe}, \citenamefont {Bourgund}, \citenamefont {Chalopin},
  \citenamefont {Bibo}, \citenamefont {Koepsell}, \citenamefont {Bojovi{\'c}},
  \citenamefont {Verresen}, \citenamefont {Pollmann}, \citenamefont {Salomon},
  \citenamefont {Gross}, \citenamefont {Hilker},\ and\ \citenamefont
  {Bloch}}]{Sompet2022}%
  \BibitemOpen
  \bibfield  {author} {\bibinfo {author} {\bibfnamefont {P.}~\bibnamefont
  {Sompet}}, \bibinfo {author} {\bibfnamefont {S.}~\bibnamefont {Hirthe}},
  \bibinfo {author} {\bibfnamefont {D.}~\bibnamefont {Bourgund}}, \bibinfo
  {author} {\bibfnamefont {T.}~\bibnamefont {Chalopin}}, \bibinfo {author}
  {\bibfnamefont {J.}~\bibnamefont {Bibo}}, \bibinfo {author} {\bibfnamefont
  {J.}~\bibnamefont {Koepsell}}, \bibinfo {author} {\bibfnamefont
  {P.}~\bibnamefont {Bojovi{\'c}}}, \bibinfo {author} {\bibfnamefont
  {R.}~\bibnamefont {Verresen}}, \bibinfo {author} {\bibfnamefont
  {F.}~\bibnamefont {Pollmann}}, \bibinfo {author} {\bibfnamefont
  {G.}~\bibnamefont {Salomon}}, \bibinfo {author} {\bibfnamefont
  {C.}~\bibnamefont {Gross}}, \bibinfo {author} {\bibfnamefont {T.~A.}\
  \bibnamefont {Hilker}},\ and\ \bibinfo {author} {\bibfnamefont
  {I.}~\bibnamefont {Bloch}},\ }\bibfield  {title} {\bibinfo {title} {Realizing
  the symmetry-protected haldane phase in fermi--hubbard ladders},\ }\href
  {https://doi.org/10.1038/s41586-022-04688-z} {\bibfield  {journal} {\bibinfo
  {journal} {Nature}\ }\textbf {\bibinfo {volume} {606}},\ \bibinfo {pages}
  {484} (\bibinfo {year} {2022})}\BibitemShut {NoStop}%
\bibitem [{\citenamefont {Wienand}\ \emph {et~al.}(2024)\citenamefont
  {Wienand}, \citenamefont {Karch}, \citenamefont {Impertro}, \citenamefont
  {Schweizer}, \citenamefont {McCulloch}, \citenamefont {Vasseur},
  \citenamefont {Gopalakrishnan}, \citenamefont {Aidelsburger},\ and\
  \citenamefont {Bloch}}]{Wienand2023}%
  \BibitemOpen
  \bibfield  {author} {\bibinfo {author} {\bibfnamefont {J.~F.}\ \bibnamefont
  {Wienand}}, \bibinfo {author} {\bibfnamefont {S.}~\bibnamefont {Karch}},
  \bibinfo {author} {\bibfnamefont {A.}~\bibnamefont {Impertro}}, \bibinfo
  {author} {\bibfnamefont {C.}~\bibnamefont {Schweizer}}, \bibinfo {author}
  {\bibfnamefont {E.}~\bibnamefont {McCulloch}}, \bibinfo {author}
  {\bibfnamefont {R.}~\bibnamefont {Vasseur}}, \bibinfo {author} {\bibfnamefont
  {S.}~\bibnamefont {Gopalakrishnan}}, \bibinfo {author} {\bibfnamefont
  {M.}~\bibnamefont {Aidelsburger}},\ and\ \bibinfo {author} {\bibfnamefont
  {I.}~\bibnamefont {Bloch}},\ }\bibfield  {title} {\bibinfo {title} {Emergence
  of fluctuating hydrodynamics in chaotic quantum systems},\ }\href
  {https://doi.org/10.1038/s41567-024-02611-z} {\bibfield  {journal} {\bibinfo
  {journal} {Nature Physics}\ }\textbf {\bibinfo {volume} {20}},\ \bibinfo
  {pages} {1732} (\bibinfo {year} {2024})}\BibitemShut {NoStop}%
\bibitem [{\citenamefont {de~Paz}\ \emph {et~al.}(2013)\citenamefont {de~Paz},
  \citenamefont {Sharma}, \citenamefont {Chotia}, \citenamefont {Mar\'echal},
  \citenamefont {Huckans}, \citenamefont {Pedri}, \citenamefont {Santos},
  \citenamefont {Gorceix}, \citenamefont {Vernac},\ and\ \citenamefont
  {Laburthe-Tolra}}]{DePaz2013}%
  \BibitemOpen
  \bibfield  {author} {\bibinfo {author} {\bibfnamefont {A.}~\bibnamefont
  {de~Paz}}, \bibinfo {author} {\bibfnamefont {A.}~\bibnamefont {Sharma}},
  \bibinfo {author} {\bibfnamefont {A.}~\bibnamefont {Chotia}}, \bibinfo
  {author} {\bibfnamefont {E.}~\bibnamefont {Mar\'echal}}, \bibinfo {author}
  {\bibfnamefont {J.~H.}\ \bibnamefont {Huckans}}, \bibinfo {author}
  {\bibfnamefont {P.}~\bibnamefont {Pedri}}, \bibinfo {author} {\bibfnamefont
  {L.}~\bibnamefont {Santos}}, \bibinfo {author} {\bibfnamefont
  {O.}~\bibnamefont {Gorceix}}, \bibinfo {author} {\bibfnamefont
  {L.}~\bibnamefont {Vernac}},\ and\ \bibinfo {author} {\bibfnamefont
  {B.}~\bibnamefont {Laburthe-Tolra}},\ }\bibfield  {title} {\bibinfo {title}
  {Nonequilibrium quantum magnetism in a dipolar lattice gas},\ }\href
  {https://doi.org/10.1103/PhysRevLett.111.185305} {\bibfield  {journal}
  {\bibinfo  {journal} {Phys. Rev. Lett.}\ }\textbf {\bibinfo {volume} {111}},\
  \bibinfo {pages} {185305} (\bibinfo {year} {2013})}\BibitemShut {NoStop}%
\bibitem [{\citenamefont {Baier}\ \emph {et~al.}(2016)\citenamefont {Baier},
  \citenamefont {Mark}, \citenamefont {Petter}, \citenamefont {Aikawa},
  \citenamefont {Chomaz}, \citenamefont {Cai}, \citenamefont {Baranov},
  \citenamefont {Zoller},\ and\ \citenamefont {Ferlaino}}]{Baier2016}%
  \BibitemOpen
  \bibfield  {author} {\bibinfo {author} {\bibfnamefont {S.}~\bibnamefont
  {Baier}}, \bibinfo {author} {\bibfnamefont {M.~J.}\ \bibnamefont {Mark}},
  \bibinfo {author} {\bibfnamefont {D.}~\bibnamefont {Petter}}, \bibinfo
  {author} {\bibfnamefont {K.}~\bibnamefont {Aikawa}}, \bibinfo {author}
  {\bibfnamefont {L.}~\bibnamefont {Chomaz}}, \bibinfo {author} {\bibfnamefont
  {Z.}~\bibnamefont {Cai}}, \bibinfo {author} {\bibfnamefont {M.}~\bibnamefont
  {Baranov}}, \bibinfo {author} {\bibfnamefont {P.}~\bibnamefont {Zoller}},\
  and\ \bibinfo {author} {\bibfnamefont {F.}~\bibnamefont {Ferlaino}},\
  }\bibfield  {title} {\bibinfo {title} {Extended bose-hubbard models with
  ultracold magnetic atoms},\ }\href {https://doi.org/10.1126/science.aac9812}
  {\bibfield  {journal} {\bibinfo  {journal} {Science}\ }\textbf {\bibinfo
  {volume} {352}},\ \bibinfo {pages} {201} (\bibinfo {year}
  {2016})}\BibitemShut {NoStop}%
\bibitem [{\citenamefont {Fersterer}\ \emph {et~al.}(2019)\citenamefont
  {Fersterer}, \citenamefont {Safavi-Naini}, \citenamefont {Zhu}, \citenamefont
  {Gabardos}, \citenamefont {Lepoutre}, \citenamefont {Vernac}, \citenamefont
  {Laburthe-Tolra}, \citenamefont {Blakie},\ and\ \citenamefont
  {Rey}}]{Fersterer2019}%
  \BibitemOpen
  \bibfield  {author} {\bibinfo {author} {\bibfnamefont {P.}~\bibnamefont
  {Fersterer}}, \bibinfo {author} {\bibfnamefont {A.}~\bibnamefont
  {Safavi-Naini}}, \bibinfo {author} {\bibfnamefont {B.}~\bibnamefont {Zhu}},
  \bibinfo {author} {\bibfnamefont {L.}~\bibnamefont {Gabardos}}, \bibinfo
  {author} {\bibfnamefont {S.}~\bibnamefont {Lepoutre}}, \bibinfo {author}
  {\bibfnamefont {L.}~\bibnamefont {Vernac}}, \bibinfo {author} {\bibfnamefont
  {B.}~\bibnamefont {Laburthe-Tolra}}, \bibinfo {author} {\bibfnamefont
  {P.~B.}\ \bibnamefont {Blakie}},\ and\ \bibinfo {author} {\bibfnamefont
  {A.~M.}\ \bibnamefont {Rey}},\ }\bibfield  {title} {\bibinfo {title}
  {Dynamics of an itinerant spin-3 atomic dipolar gas in an optical lattice},\
  }\href {https://doi.org/10.1103/PhysRevA.100.033609} {\bibfield  {journal}
  {\bibinfo  {journal} {Phys. Rev. A}\ }\textbf {\bibinfo {volume} {100}},\
  \bibinfo {pages} {033609} (\bibinfo {year} {2019})}\BibitemShut {NoStop}%
\bibitem [{\citenamefont {Su}\ \emph {et~al.}(2023)\citenamefont {Su},
  \citenamefont {Douglas}, \citenamefont {Szurek}, \citenamefont {Groth},
  \citenamefont {Ozturk}, \citenamefont {Krahn}, \citenamefont {H{\'e}bert},
  \citenamefont {Phelps}, \citenamefont {Ebadi}, \citenamefont {Dickerson},
  \citenamefont {Ferlaino}, \citenamefont {Markovi{\'c}},\ and\ \citenamefont
  {Greiner}}]{Su2023}%
  \BibitemOpen
  \bibfield  {author} {\bibinfo {author} {\bibfnamefont {L.}~\bibnamefont
  {Su}}, \bibinfo {author} {\bibfnamefont {A.}~\bibnamefont {Douglas}},
  \bibinfo {author} {\bibfnamefont {M.}~\bibnamefont {Szurek}}, \bibinfo
  {author} {\bibfnamefont {R.}~\bibnamefont {Groth}}, \bibinfo {author}
  {\bibfnamefont {S.~F.}\ \bibnamefont {Ozturk}}, \bibinfo {author}
  {\bibfnamefont {A.}~\bibnamefont {Krahn}}, \bibinfo {author} {\bibfnamefont
  {A.~H.}\ \bibnamefont {H{\'e}bert}}, \bibinfo {author} {\bibfnamefont
  {G.~A.}\ \bibnamefont {Phelps}}, \bibinfo {author} {\bibfnamefont
  {S.}~\bibnamefont {Ebadi}}, \bibinfo {author} {\bibfnamefont
  {S.}~\bibnamefont {Dickerson}}, \bibinfo {author} {\bibfnamefont
  {F.}~\bibnamefont {Ferlaino}}, \bibinfo {author} {\bibfnamefont
  {O.}~\bibnamefont {Markovi{\'c}}},\ and\ \bibinfo {author} {\bibfnamefont
  {M.}~\bibnamefont {Greiner}},\ }\bibfield  {title} {\bibinfo {title} {Dipolar
  quantum solids emerging in a hubbard quantum simulator},\ }\href
  {https://doi.org/10.1038/s41586-023-06614-3} {\bibfield  {journal} {\bibinfo
  {journal} {Nature}\ }\textbf {\bibinfo {volume} {622}},\ \bibinfo {pages}
  {724} (\bibinfo {year} {2023})}\BibitemShut {NoStop}%
\bibitem [{\citenamefont {Yan}\ \emph {et~al.}(2013)\citenamefont {Yan},
  \citenamefont {Moses}, \citenamefont {Gadway}, \citenamefont {Covey},
  \citenamefont {Hazzard}, \citenamefont {Rey}, \citenamefont {Jin},\ and\
  \citenamefont {Ye}}]{Yan2013}%
  \BibitemOpen
  \bibfield  {author} {\bibinfo {author} {\bibfnamefont {B.}~\bibnamefont
  {Yan}}, \bibinfo {author} {\bibfnamefont {S.~A.}\ \bibnamefont {Moses}},
  \bibinfo {author} {\bibfnamefont {B.}~\bibnamefont {Gadway}}, \bibinfo
  {author} {\bibfnamefont {J.~P.}\ \bibnamefont {Covey}}, \bibinfo {author}
  {\bibfnamefont {K.~R.~A.}\ \bibnamefont {Hazzard}}, \bibinfo {author}
  {\bibfnamefont {A.~M.}\ \bibnamefont {Rey}}, \bibinfo {author} {\bibfnamefont
  {D.~S.}\ \bibnamefont {Jin}},\ and\ \bibinfo {author} {\bibfnamefont
  {J.}~\bibnamefont {Ye}},\ }\bibfield  {title} {\bibinfo {title} {Observation
  of dipolar spin-exchange interactions with lattice-confined polar
  molecules},\ }\href {https://doi.org/10.1038/nature12483} {\bibfield
  {journal} {\bibinfo  {journal} {Nature}\ }\textbf {\bibinfo {volume} {501}},\
  \bibinfo {pages} {521} (\bibinfo {year} {2013})}\BibitemShut {NoStop}%
\bibitem [{\citenamefont {Christakis}\ \emph {et~al.}(2023)\citenamefont
  {Christakis}, \citenamefont {Rosenberg}, \citenamefont {Raj}, \citenamefont
  {Chi}, \citenamefont {Morningstar}, \citenamefont {Huse}, \citenamefont
  {Yan},\ and\ \citenamefont {Bakr}}]{Christakis2023}%
  \BibitemOpen
  \bibfield  {author} {\bibinfo {author} {\bibfnamefont {L.}~\bibnamefont
  {Christakis}}, \bibinfo {author} {\bibfnamefont {J.~S.}\ \bibnamefont
  {Rosenberg}}, \bibinfo {author} {\bibfnamefont {R.}~\bibnamefont {Raj}},
  \bibinfo {author} {\bibfnamefont {S.}~\bibnamefont {Chi}}, \bibinfo {author}
  {\bibfnamefont {A.}~\bibnamefont {Morningstar}}, \bibinfo {author}
  {\bibfnamefont {D.~A.}\ \bibnamefont {Huse}}, \bibinfo {author}
  {\bibfnamefont {Z.~Z.}\ \bibnamefont {Yan}},\ and\ \bibinfo {author}
  {\bibfnamefont {W.~S.}\ \bibnamefont {Bakr}},\ }\bibfield  {title} {\bibinfo
  {title} {Probing site-resolved correlations in a spin system of ultracold
  molecules},\ }\href {https://doi.org/10.1038/s41586-022-05558-4} {\bibfield
  {journal} {\bibinfo  {journal} {Nature}\ }\textbf {\bibinfo {volume} {614}},\
  \bibinfo {pages} {64} (\bibinfo {year} {2023})}\BibitemShut {NoStop}%
\bibitem [{\citenamefont {Holland}\ \emph {et~al.}(2023)\citenamefont
  {Holland}, \citenamefont {Lu},\ and\ \citenamefont {Cheuk}}]{Holland2023}%
  \BibitemOpen
  \bibfield  {author} {\bibinfo {author} {\bibfnamefont {C.~M.}\ \bibnamefont
  {Holland}}, \bibinfo {author} {\bibfnamefont {Y.}~\bibnamefont {Lu}},\ and\
  \bibinfo {author} {\bibfnamefont {L.~W.}\ \bibnamefont {Cheuk}},\ }\bibfield
  {title} {\bibinfo {title} {On-demand entanglement of molecules in a
  reconfigurable optical tweezer array},\ }\href
  {https://doi.org/10.1126/science.adf4272} {\bibfield  {journal} {\bibinfo
  {journal} {Science}\ }\textbf {\bibinfo {volume} {382}},\ \bibinfo {pages}
  {1143} (\bibinfo {year} {2023})}\BibitemShut {NoStop}%
\bibitem [{\citenamefont {Bao}\ \emph {et~al.}(2023)\citenamefont {Bao},
  \citenamefont {Yu}, \citenamefont {Anderegg}, \citenamefont {Chae},
  \citenamefont {Ketterle}, \citenamefont {Ni},\ and\ \citenamefont
  {Doyle}}]{Bao2023}%
  \BibitemOpen
  \bibfield  {author} {\bibinfo {author} {\bibfnamefont {Y.}~\bibnamefont
  {Bao}}, \bibinfo {author} {\bibfnamefont {S.~S.}\ \bibnamefont {Yu}},
  \bibinfo {author} {\bibfnamefont {L.}~\bibnamefont {Anderegg}}, \bibinfo
  {author} {\bibfnamefont {E.}~\bibnamefont {Chae}}, \bibinfo {author}
  {\bibfnamefont {W.}~\bibnamefont {Ketterle}}, \bibinfo {author}
  {\bibfnamefont {K.-K.}\ \bibnamefont {Ni}},\ and\ \bibinfo {author}
  {\bibfnamefont {J.~M.}\ \bibnamefont {Doyle}},\ }\bibfield  {title} {\bibinfo
  {title} {Dipolar spin-exchange and entanglement between molecules in an
  optical tweezer array},\ }\href {https://doi.org/10.1126/science.adf8999}
  {\bibfield  {journal} {\bibinfo  {journal} {Science}\ }\textbf {\bibinfo
  {volume} {382}},\ \bibinfo {pages} {1138} (\bibinfo {year}
  {2023})}\BibitemShut {NoStop}%
\bibitem [{\citenamefont {Carroll}\ \emph {et~al.}(2024)\citenamefont
  {Carroll}, \citenamefont {Hirzler}, \citenamefont {Miller}, \citenamefont
  {Wellnitz}, \citenamefont {Muleady}, \citenamefont {Lin}, \citenamefont
  {Zamarski}, \citenamefont {Wang}, \citenamefont {Bohn}, \citenamefont {Rey},\
  and\ \citenamefont {Ye}}]{Carroll2024}%
  \BibitemOpen
  \bibfield  {author} {\bibinfo {author} {\bibfnamefont {A.~N.}\ \bibnamefont
  {Carroll}}, \bibinfo {author} {\bibfnamefont {H.}~\bibnamefont {Hirzler}},
  \bibinfo {author} {\bibfnamefont {C.}~\bibnamefont {Miller}}, \bibinfo
  {author} {\bibfnamefont {D.}~\bibnamefont {Wellnitz}}, \bibinfo {author}
  {\bibfnamefont {S.~R.}\ \bibnamefont {Muleady}}, \bibinfo {author}
  {\bibfnamefont {J.}~\bibnamefont {Lin}}, \bibinfo {author} {\bibfnamefont
  {K.~P.}\ \bibnamefont {Zamarski}}, \bibinfo {author} {\bibfnamefont
  {R.~R.~W.}\ \bibnamefont {Wang}}, \bibinfo {author} {\bibfnamefont {J.~L.}\
  \bibnamefont {Bohn}}, \bibinfo {author} {\bibfnamefont {A.~M.}\ \bibnamefont
  {Rey}},\ and\ \bibinfo {author} {\bibfnamefont {J.}~\bibnamefont {Ye}},\
  }\href@noop {} {\bibinfo {title} {Observation of generalized t-j spin
  dynamics with tunable dipolar interactions}} (\bibinfo {year} {2024}),\
  \Eprint {https://arxiv.org/abs/2404.18916} {arXiv:2404.18916
  [cond-mat.quant-gas]} \BibitemShut {NoStop}%
\bibitem [{\citenamefont {de~L\'es\'eleuc}\ \emph {et~al.}(2019)\citenamefont
  {de~L\'es\'eleuc}, \citenamefont {Lienhard}, \citenamefont {Scholl},
  \citenamefont {Barredo}, \citenamefont {Weber}, \citenamefont {Lang},
  \citenamefont {B\"uchler}, \citenamefont {Lahaye},\ and\ \citenamefont
  {Browaeys}}]{Leseluc2019}%
  \BibitemOpen
  \bibfield  {author} {\bibinfo {author} {\bibfnamefont {S.}~\bibnamefont
  {de~L\'es\'eleuc}}, \bibinfo {author} {\bibfnamefont {V.}~\bibnamefont
  {Lienhard}}, \bibinfo {author} {\bibfnamefont {P.}~\bibnamefont {Scholl}},
  \bibinfo {author} {\bibfnamefont {D.}~\bibnamefont {Barredo}}, \bibinfo
  {author} {\bibfnamefont {S.}~\bibnamefont {Weber}}, \bibinfo {author}
  {\bibfnamefont {N.}~\bibnamefont {Lang}}, \bibinfo {author} {\bibfnamefont
  {H.~P.}\ \bibnamefont {B\"uchler}}, \bibinfo {author} {\bibfnamefont
  {T.}~\bibnamefont {Lahaye}},\ and\ \bibinfo {author} {\bibfnamefont
  {A.}~\bibnamefont {Browaeys}},\ }\bibfield  {title} {\bibinfo {title}
  {Observation of a symmetry-protected topological phase of interacting bosons
  with rydberg atoms},\ }\href {https://doi.org/10.1126/science.aav9105}
  {\bibfield  {journal} {\bibinfo  {journal} {Science}\ }\textbf {\bibinfo
  {volume} {365}},\ \bibinfo {pages} {775} (\bibinfo {year}
  {2019})}\BibitemShut {NoStop}%
\bibitem [{\citenamefont {Lienhard}\ \emph {et~al.}(2020)\citenamefont
  {Lienhard}, \citenamefont {Scholl}, \citenamefont {Weber}, \citenamefont
  {Barredo}, \citenamefont {de~L\'es\'eleuc}, \citenamefont {Bai},
  \citenamefont {Lang}, \citenamefont {Fleischhauer}, \citenamefont
  {B\"uchler}, \citenamefont {Lahaye},\ and\ \citenamefont
  {Browaeys}}]{Lienhard2020}%
  \BibitemOpen
  \bibfield  {author} {\bibinfo {author} {\bibfnamefont {V.}~\bibnamefont
  {Lienhard}}, \bibinfo {author} {\bibfnamefont {P.}~\bibnamefont {Scholl}},
  \bibinfo {author} {\bibfnamefont {S.}~\bibnamefont {Weber}}, \bibinfo
  {author} {\bibfnamefont {D.}~\bibnamefont {Barredo}}, \bibinfo {author}
  {\bibfnamefont {S.}~\bibnamefont {de~L\'es\'eleuc}}, \bibinfo {author}
  {\bibfnamefont {R.}~\bibnamefont {Bai}}, \bibinfo {author} {\bibfnamefont
  {N.}~\bibnamefont {Lang}}, \bibinfo {author} {\bibfnamefont {M.}~\bibnamefont
  {Fleischhauer}}, \bibinfo {author} {\bibfnamefont {H.~P.}\ \bibnamefont
  {B\"uchler}}, \bibinfo {author} {\bibfnamefont {T.}~\bibnamefont {Lahaye}},\
  and\ \bibinfo {author} {\bibfnamefont {A.}~\bibnamefont {Browaeys}},\
  }\bibfield  {title} {\bibinfo {title} {Realization of a density-dependent
  peierls phase in a synthetic, spin-orbit coupled rydberg system},\ }\href
  {https://doi.org/10.1103/PhysRevX.10.021031} {\bibfield  {journal} {\bibinfo
  {journal} {Phys. Rev. X}\ }\textbf {\bibinfo {volume} {10}},\ \bibinfo
  {pages} {021031} (\bibinfo {year} {2020})}\BibitemShut {NoStop}%
\bibitem [{\citenamefont {Signoles}\ \emph {et~al.}(2021)\citenamefont
  {Signoles}, \citenamefont {Franz}, \citenamefont {Ferracini~Alves},
  \citenamefont {G\"arttner}, \citenamefont {Whitlock}, \citenamefont
  {Z\"urn},\ and\ \citenamefont {Weidem\"uller}}]{Signoles2021}%
  \BibitemOpen
  \bibfield  {author} {\bibinfo {author} {\bibfnamefont {A.}~\bibnamefont
  {Signoles}}, \bibinfo {author} {\bibfnamefont {T.}~\bibnamefont {Franz}},
  \bibinfo {author} {\bibfnamefont {R.}~\bibnamefont {Ferracini~Alves}},
  \bibinfo {author} {\bibfnamefont {M.}~\bibnamefont {G\"arttner}}, \bibinfo
  {author} {\bibfnamefont {S.}~\bibnamefont {Whitlock}}, \bibinfo {author}
  {\bibfnamefont {G.}~\bibnamefont {Z\"urn}},\ and\ \bibinfo {author}
  {\bibfnamefont {M.}~\bibnamefont {Weidem\"uller}},\ }\bibfield  {title}
  {\bibinfo {title} {Glassy dynamics in a disordered heisenberg quantum spin
  system},\ }\href {https://doi.org/10.1103/PhysRevX.11.011011} {\bibfield
  {journal} {\bibinfo  {journal} {Phys. Rev. X}\ }\textbf {\bibinfo {volume}
  {11}},\ \bibinfo {pages} {011011} (\bibinfo {year} {2021})}\BibitemShut
  {NoStop}%
\bibitem [{\citenamefont {Scholl}\ \emph {et~al.}(2022)\citenamefont {Scholl},
  \citenamefont {Williams}, \citenamefont {Bornet}, \citenamefont {Wallner},
  \citenamefont {Barredo}, \citenamefont {Henriet}, \citenamefont {Signoles},
  \citenamefont {Hainaut}, \citenamefont {Franz}, \citenamefont {Geier},
  \citenamefont {Tebben}, \citenamefont {Salzinger}, \citenamefont {Z\"urn},
  \citenamefont {Lahaye}, \citenamefont {Weidem\"uller},\ and\ \citenamefont
  {Browaeys}}]{Scholl2022}%
  \BibitemOpen
  \bibfield  {author} {\bibinfo {author} {\bibfnamefont {P.}~\bibnamefont
  {Scholl}}, \bibinfo {author} {\bibfnamefont {H.~J.}\ \bibnamefont
  {Williams}}, \bibinfo {author} {\bibfnamefont {G.}~\bibnamefont {Bornet}},
  \bibinfo {author} {\bibfnamefont {F.}~\bibnamefont {Wallner}}, \bibinfo
  {author} {\bibfnamefont {D.}~\bibnamefont {Barredo}}, \bibinfo {author}
  {\bibfnamefont {L.}~\bibnamefont {Henriet}}, \bibinfo {author} {\bibfnamefont
  {A.}~\bibnamefont {Signoles}}, \bibinfo {author} {\bibfnamefont
  {C.}~\bibnamefont {Hainaut}}, \bibinfo {author} {\bibfnamefont
  {T.}~\bibnamefont {Franz}}, \bibinfo {author} {\bibfnamefont
  {S.}~\bibnamefont {Geier}}, \bibinfo {author} {\bibfnamefont
  {A.}~\bibnamefont {Tebben}}, \bibinfo {author} {\bibfnamefont
  {A.}~\bibnamefont {Salzinger}}, \bibinfo {author} {\bibfnamefont
  {G.}~\bibnamefont {Z\"urn}}, \bibinfo {author} {\bibfnamefont
  {T.}~\bibnamefont {Lahaye}}, \bibinfo {author} {\bibfnamefont
  {M.}~\bibnamefont {Weidem\"uller}},\ and\ \bibinfo {author} {\bibfnamefont
  {A.}~\bibnamefont {Browaeys}},\ }\bibfield  {title} {\bibinfo {title}
  {Microwave engineering of programmable $xxz$ hamiltonians in arrays of
  rydberg atoms},\ }\href {https://doi.org/10.1103/PRXQuantum.3.020303}
  {\bibfield  {journal} {\bibinfo  {journal} {PRX Quantum}\ }\textbf {\bibinfo
  {volume} {3}},\ \bibinfo {pages} {020303} (\bibinfo {year}
  {2022})}\BibitemShut {NoStop}%
\bibitem [{\citenamefont {Bornet}\ \emph {et~al.}(2023)\citenamefont {Bornet},
  \citenamefont {Emperauger}, \citenamefont {Chen}, \citenamefont {Ye},
  \citenamefont {Block}, \citenamefont {Bintz}, \citenamefont {Boyd},
  \citenamefont {Barredo}, \citenamefont {Comparin}, \citenamefont {Mezzacapo},
  \citenamefont {Roscilde}, \citenamefont {Lahaye}, \citenamefont {Yao},\ and\
  \citenamefont {Browaeys}}]{Bornet2023}%
  \BibitemOpen
  \bibfield  {author} {\bibinfo {author} {\bibfnamefont {G.}~\bibnamefont
  {Bornet}}, \bibinfo {author} {\bibfnamefont {G.}~\bibnamefont {Emperauger}},
  \bibinfo {author} {\bibfnamefont {C.}~\bibnamefont {Chen}}, \bibinfo {author}
  {\bibfnamefont {B.}~\bibnamefont {Ye}}, \bibinfo {author} {\bibfnamefont
  {M.}~\bibnamefont {Block}}, \bibinfo {author} {\bibfnamefont
  {M.}~\bibnamefont {Bintz}}, \bibinfo {author} {\bibfnamefont {J.~A.}\
  \bibnamefont {Boyd}}, \bibinfo {author} {\bibfnamefont {D.}~\bibnamefont
  {Barredo}}, \bibinfo {author} {\bibfnamefont {T.}~\bibnamefont {Comparin}},
  \bibinfo {author} {\bibfnamefont {F.}~\bibnamefont {Mezzacapo}}, \bibinfo
  {author} {\bibfnamefont {T.}~\bibnamefont {Roscilde}}, \bibinfo {author}
  {\bibfnamefont {T.}~\bibnamefont {Lahaye}}, \bibinfo {author} {\bibfnamefont
  {N.~Y.}\ \bibnamefont {Yao}},\ and\ \bibinfo {author} {\bibfnamefont
  {A.}~\bibnamefont {Browaeys}},\ }\bibfield  {title} {\bibinfo {title}
  {Scalable spin squeezing in a dipolar rydberg atom array},\ }\href
  {https://doi.org/10.1038/s41586-023-06414-9} {\bibfield  {journal} {\bibinfo
  {journal} {Nature}\ }\textbf {\bibinfo {volume} {621}},\ \bibinfo {pages}
  {728} (\bibinfo {year} {2023})}\BibitemShut {NoStop}%
\bibitem [{\citenamefont {Chanda}\ \emph {et~al.}(2024)\citenamefont {Chanda},
  \citenamefont {Barbiero}, \citenamefont {Lewenstein}, \citenamefont {Mark},\
  and\ \citenamefont {Zakrzewski}}]{Chanda2024}%
  \BibitemOpen
  \bibfield  {author} {\bibinfo {author} {\bibfnamefont {T.}~\bibnamefont
  {Chanda}}, \bibinfo {author} {\bibfnamefont {L.}~\bibnamefont {Barbiero}},
  \bibinfo {author} {\bibfnamefont {M.}~\bibnamefont {Lewenstein}}, \bibinfo
  {author} {\bibfnamefont {M.~J.}\ \bibnamefont {Mark}},\ and\ \bibinfo
  {author} {\bibfnamefont {J.}~\bibnamefont {Zakrzewski}},\ }\href@noop {}
  {\bibinfo {title} {Recent progress on quantum simulations of non-standard
  bose-hubbard models}} (\bibinfo {year} {2024}),\ \Eprint
  {https://arxiv.org/abs/2405.07775} {arXiv:2405.07775} \BibitemShut {NoStop}%
\bibitem [{\citenamefont {Pokrovsky}\ and\ \citenamefont
  {Talapov}(1979)}]{Pokrovsky1979}%
  \BibitemOpen
  \bibfield  {author} {\bibinfo {author} {\bibfnamefont {V.~L.}\ \bibnamefont
  {Pokrovsky}}\ and\ \bibinfo {author} {\bibfnamefont {A.~L.}\ \bibnamefont
  {Talapov}},\ }\bibfield  {title} {\bibinfo {title} {Ground state, spectrum,
  and phase diagram of two-dimensional incommensurate crystals},\ }\href
  {https://doi.org/10.1103/PhysRevLett.42.65} {\bibfield  {journal} {\bibinfo
  {journal} {Phys. Rev. Lett.}\ }\textbf {\bibinfo {volume} {42}},\ \bibinfo
  {pages} {65} (\bibinfo {year} {1979})}\BibitemShut {NoStop}%
\bibitem [{\citenamefont {Bak}(1982)}]{Bak1982}%
  \BibitemOpen
  \bibfield  {author} {\bibinfo {author} {\bibfnamefont {P.}~\bibnamefont
  {Bak}},\ }\bibfield  {title} {\bibinfo {title} {Commensurate phases,
  incommensurate phases and the devil's staircase},\ }\href
  {https://doi.org/10.1088/0034-4885/45/6/001} {\bibfield  {journal} {\bibinfo
  {journal} {Reports on Progress in Physics}\ }\textbf {\bibinfo {volume}
  {45}},\ \bibinfo {pages} {587} (\bibinfo {year} {1982})}\BibitemShut
  {NoStop}%
\bibitem [{\citenamefont {Rieger}\ and\ \citenamefont
  {Uimin}(1996)}]{Rieger1996}%
  \BibitemOpen
  \bibfield  {author} {\bibinfo {author} {\bibfnamefont {H.}~\bibnamefont
  {Rieger}}\ and\ \bibinfo {author} {\bibfnamefont {G.}~\bibnamefont {Uimin}},\
  }\bibfield  {title} {\bibinfo {title} {The one-dimensional annni model in a
  transverse field: analytic and numerical study of effective hamiltonians},\
  }\href {https://doi.org/10.1007/s002570050252} {\bibfield  {journal}
  {\bibinfo  {journal} {Zeitschrift f{\"u}r Physik B Condensed Matter}\
  }\textbf {\bibinfo {volume} {101}},\ \bibinfo {pages} {597} (\bibinfo {year}
  {1996})}\BibitemShut {NoStop}%
\bibitem [{\citenamefont {Fendley}\ \emph {et~al.}(2004)\citenamefont
  {Fendley}, \citenamefont {Sengupta},\ and\ \citenamefont
  {Sachdev}}]{Fendley2004}%
  \BibitemOpen
  \bibfield  {author} {\bibinfo {author} {\bibfnamefont {P.}~\bibnamefont
  {Fendley}}, \bibinfo {author} {\bibfnamefont {K.}~\bibnamefont {Sengupta}},\
  and\ \bibinfo {author} {\bibfnamefont {S.}~\bibnamefont {Sachdev}},\
  }\bibfield  {title} {\bibinfo {title} {Competing density-wave orders in a
  one-dimensional hard-boson model},\ }\href
  {https://doi.org/10.1103/PhysRevB.69.075106} {\bibfield  {journal} {\bibinfo
  {journal} {Phys. Rev. B}\ }\textbf {\bibinfo {volume} {69}},\ \bibinfo
  {pages} {075106} (\bibinfo {year} {2004})}\BibitemShut {NoStop}%
\bibitem [{\citenamefont {Weimer}\ and\ \citenamefont
  {B\"uchler}(2010)}]{Weimer2010}%
  \BibitemOpen
  \bibfield  {author} {\bibinfo {author} {\bibfnamefont {H.}~\bibnamefont
  {Weimer}}\ and\ \bibinfo {author} {\bibfnamefont {H.~P.}\ \bibnamefont
  {B\"uchler}},\ }\bibfield  {title} {\bibinfo {title} {Two-stage melting in
  systems of strongly interacting rydberg atoms},\ }\href
  {https://doi.org/10.1103/PhysRevLett.105.230403} {\bibfield  {journal}
  {\bibinfo  {journal} {Phys. Rev. Lett.}\ }\textbf {\bibinfo {volume} {105}},\
  \bibinfo {pages} {230403} (\bibinfo {year} {2010})}\BibitemShut {NoStop}%
\bibitem [{\citenamefont {Bermudez}\ \emph {et~al.}(2012)\citenamefont
  {Bermudez}, \citenamefont {Almeida}, \citenamefont {Ott}, \citenamefont
  {Kaufmann}, \citenamefont {Ulm}, \citenamefont {Poschinger}, \citenamefont
  {Schmidt-Kaler}, \citenamefont {Retzker},\ and\ \citenamefont
  {Plenio}}]{Bermudez2012}%
  \BibitemOpen
  \bibfield  {author} {\bibinfo {author} {\bibfnamefont {A.}~\bibnamefont
  {Bermudez}}, \bibinfo {author} {\bibfnamefont {J.}~\bibnamefont {Almeida}},
  \bibinfo {author} {\bibfnamefont {K.}~\bibnamefont {Ott}}, \bibinfo {author}
  {\bibfnamefont {H.}~\bibnamefont {Kaufmann}}, \bibinfo {author}
  {\bibfnamefont {S.}~\bibnamefont {Ulm}}, \bibinfo {author} {\bibfnamefont
  {U.}~\bibnamefont {Poschinger}}, \bibinfo {author} {\bibfnamefont
  {F.}~\bibnamefont {Schmidt-Kaler}}, \bibinfo {author} {\bibfnamefont
  {A.}~\bibnamefont {Retzker}},\ and\ \bibinfo {author} {\bibfnamefont {M.~B.}\
  \bibnamefont {Plenio}},\ }\bibfield  {title} {\bibinfo {title} {Quantum
  magnetism of spin-ladder compounds with trapped-ion crystals},\ }\href
  {https://doi.org/10.1088/1367-2630/14/9/093042} {\bibfield  {journal}
  {\bibinfo  {journal} {New Journal of Physics}\ }\textbf {\bibinfo {volume}
  {14}},\ \bibinfo {pages} {093042} (\bibinfo {year} {2012})}\BibitemShut
  {NoStop}%
\bibitem [{\citenamefont {Weimer}(2014)}]{Weimer2014}%
  \BibitemOpen
  \bibfield  {author} {\bibinfo {author} {\bibfnamefont {H.}~\bibnamefont
  {Weimer}},\ }\bibfield  {title} {\bibinfo {title} {String order in
  dipole-blockaded quantum liquids},\ }\href
  {https://doi.org/10.1088/1367-2630/16/9/093040} {\bibfield  {journal}
  {\bibinfo  {journal} {New Journal of Physics}\ }\textbf {\bibinfo {volume}
  {16}},\ \bibinfo {pages} {093040} (\bibinfo {year} {2014})}\BibitemShut
  {NoStop}%
\bibitem [{\citenamefont {Chepiga}\ and\ \citenamefont
  {Mila}(2019)}]{Chepiga2019}%
  \BibitemOpen
  \bibfield  {author} {\bibinfo {author} {\bibfnamefont {N.}~\bibnamefont
  {Chepiga}}\ and\ \bibinfo {author} {\bibfnamefont {F.}~\bibnamefont {Mila}},\
  }\bibfield  {title} {\bibinfo {title} {Floating phase versus chiral
  transition in a 1d hard-boson model},\ }\href
  {https://doi.org/10.1103/PhysRevLett.122.017205} {\bibfield  {journal}
  {\bibinfo  {journal} {Phys. Rev. Lett.}\ }\textbf {\bibinfo {volume} {122}},\
  \bibinfo {pages} {017205} (\bibinfo {year} {2019})}\BibitemShut {NoStop}%
\bibitem [{\citenamefont {Giudici}\ \emph {et~al.}(2019)\citenamefont
  {Giudici}, \citenamefont {Angelone}, \citenamefont {Magnifico}, \citenamefont
  {Zeng}, \citenamefont {Giudice}, \citenamefont {Mendes-Santos},\ and\
  \citenamefont {Dalmonte}}]{Giudici2019}%
  \BibitemOpen
  \bibfield  {author} {\bibinfo {author} {\bibfnamefont {G.}~\bibnamefont
  {Giudici}}, \bibinfo {author} {\bibfnamefont {A.}~\bibnamefont {Angelone}},
  \bibinfo {author} {\bibfnamefont {G.}~\bibnamefont {Magnifico}}, \bibinfo
  {author} {\bibfnamefont {Z.}~\bibnamefont {Zeng}}, \bibinfo {author}
  {\bibfnamefont {G.}~\bibnamefont {Giudice}}, \bibinfo {author} {\bibfnamefont
  {T.}~\bibnamefont {Mendes-Santos}},\ and\ \bibinfo {author} {\bibfnamefont
  {M.}~\bibnamefont {Dalmonte}},\ }\bibfield  {title} {\bibinfo {title}
  {Diagnosing potts criticality and two-stage melting in one-dimensional
  hard-core boson models},\ }\href {https://doi.org/10.1103/PhysRevB.99.094434}
  {\bibfield  {journal} {\bibinfo  {journal} {Phys. Rev. B}\ }\textbf {\bibinfo
  {volume} {99}},\ \bibinfo {pages} {094434} (\bibinfo {year}
  {2019})}\BibitemShut {NoStop}%
\bibitem [{\citenamefont {Rader}\ and\ \citenamefont
  {Läuchli}(2019)}]{Rader2019}%
  \BibitemOpen
  \bibfield  {author} {\bibinfo {author} {\bibfnamefont {M.}~\bibnamefont
  {Rader}}\ and\ \bibinfo {author} {\bibfnamefont {A.~M.}\ \bibnamefont
  {Läuchli}},\ }\href@noop {} {\bibinfo {title} {Floating phases in
  one-dimensional rydberg ising chains}} (\bibinfo {year} {2019}),\ \Eprint
  {https://arxiv.org/abs/1908.02068} {arXiv:1908.02068 [cond-mat.quant-gas]}
  \BibitemShut {NoStop}%
\bibitem [{\citenamefont {Chepiga}\ and\ \citenamefont
  {Mila}(2021)}]{Chepiga2021}%
  \BibitemOpen
  \bibfield  {author} {\bibinfo {author} {\bibfnamefont {N.}~\bibnamefont
  {Chepiga}}\ and\ \bibinfo {author} {\bibfnamefont {F.}~\bibnamefont {Mila}},\
  }\bibfield  {title} {\bibinfo {title} {Kibble-zurek exponent and chiral
  transition of the period-4 phase of rydberg chains},\ }\href
  {https://doi.org/10.1038/s41467-020-20641-y} {\bibfield  {journal} {\bibinfo
  {journal} {Nature Communications}\ }\textbf {\bibinfo {volume} {12}},\
  \bibinfo {pages} {414} (\bibinfo {year} {2021})}\BibitemShut {NoStop}%
\bibitem [{\citenamefont {Maceira}\ \emph {et~al.}(2022)\citenamefont
  {Maceira}, \citenamefont {Chepiga},\ and\ \citenamefont
  {Mila}}]{Maceira2022}%
  \BibitemOpen
  \bibfield  {author} {\bibinfo {author} {\bibfnamefont {I.~A.}\ \bibnamefont
  {Maceira}}, \bibinfo {author} {\bibfnamefont {N.}~\bibnamefont {Chepiga}},\
  and\ \bibinfo {author} {\bibfnamefont {F.}~\bibnamefont {Mila}},\ }\bibfield
  {title} {\bibinfo {title} {Conformal and chiral phase transitions in rydberg
  chains},\ }\href {https://doi.org/10.1103/PhysRevResearch.4.043102}
  {\bibfield  {journal} {\bibinfo  {journal} {Phys. Rev. Res.}\ }\textbf
  {\bibinfo {volume} {4}},\ \bibinfo {pages} {043102} (\bibinfo {year}
  {2022})}\BibitemShut {NoStop}%
\bibitem [{\citenamefont {Zhang}\ \emph {et~al.}(2025)\citenamefont {Zhang},
  \citenamefont {Cant{\'u}}, \citenamefont {Liu}, \citenamefont {Bylinskii},
  \citenamefont {Braverman}, \citenamefont {Huber}, \citenamefont
  {Amato-Grill}, \citenamefont {Lukin}, \citenamefont {Gemelke}, \citenamefont
  {Keesling}, \citenamefont {Wang}, \citenamefont {Meurice},\ and\
  \citenamefont {Tsai}}]{Zhang2024}%
  \BibitemOpen
  \bibfield  {author} {\bibinfo {author} {\bibfnamefont {J.}~\bibnamefont
  {Zhang}}, \bibinfo {author} {\bibfnamefont {S.~H.}\ \bibnamefont
  {Cant{\'u}}}, \bibinfo {author} {\bibfnamefont {F.}~\bibnamefont {Liu}},
  \bibinfo {author} {\bibfnamefont {A.}~\bibnamefont {Bylinskii}}, \bibinfo
  {author} {\bibfnamefont {B.}~\bibnamefont {Braverman}}, \bibinfo {author}
  {\bibfnamefont {F.}~\bibnamefont {Huber}}, \bibinfo {author} {\bibfnamefont
  {J.}~\bibnamefont {Amato-Grill}}, \bibinfo {author} {\bibfnamefont
  {A.}~\bibnamefont {Lukin}}, \bibinfo {author} {\bibfnamefont
  {N.}~\bibnamefont {Gemelke}}, \bibinfo {author} {\bibfnamefont
  {A.}~\bibnamefont {Keesling}}, \bibinfo {author} {\bibfnamefont {S.-T.}\
  \bibnamefont {Wang}}, \bibinfo {author} {\bibfnamefont {Y.}~\bibnamefont
  {Meurice}},\ and\ \bibinfo {author} {\bibfnamefont {S.-W.}\ \bibnamefont
  {Tsai}},\ }\bibfield  {title} {\bibinfo {title} {Probing quantum floating
  phases in rydberg atom arrays},\ }\href
  {https://doi.org/10.1038/s41467-025-55947-2} {\bibfield  {journal} {\bibinfo
  {journal} {Nature Communications}\ }\textbf {\bibinfo {volume} {16}},\
  \bibinfo {pages} {712} (\bibinfo {year} {2025})}\BibitemShut {NoStop}%
\bibitem [{\citenamefont {Kestner}\ \emph {et~al.}(2011)\citenamefont
  {Kestner}, \citenamefont {Wang}, \citenamefont {Sau},\ and\ \citenamefont
  {Das~Sarma}}]{Kestner2011}%
  \BibitemOpen
  \bibfield  {author} {\bibinfo {author} {\bibfnamefont {J.~P.}\ \bibnamefont
  {Kestner}}, \bibinfo {author} {\bibfnamefont {B.}~\bibnamefont {Wang}},
  \bibinfo {author} {\bibfnamefont {J.~D.}\ \bibnamefont {Sau}},\ and\ \bibinfo
  {author} {\bibfnamefont {S.}~\bibnamefont {Das~Sarma}},\ }\bibfield  {title}
  {\bibinfo {title} {Prediction of a gapless topological haldane liquid phase
  in a one-dimensional cold polar molecular lattice},\ }\href
  {https://doi.org/10.1103/PhysRevB.83.174409} {\bibfield  {journal} {\bibinfo
  {journal} {Phys. Rev. B}\ }\textbf {\bibinfo {volume} {83}},\ \bibinfo
  {pages} {174409} (\bibinfo {year} {2011})}\BibitemShut {NoStop}%
\bibitem [{\citenamefont {Ruhman}\ \emph {et~al.}(2012)\citenamefont {Ruhman},
  \citenamefont {Dalla~Torre}, \citenamefont {Huber},\ and\ \citenamefont
  {Altman}}]{Ruhman2012}%
  \BibitemOpen
  \bibfield  {author} {\bibinfo {author} {\bibfnamefont {J.}~\bibnamefont
  {Ruhman}}, \bibinfo {author} {\bibfnamefont {E.~G.}\ \bibnamefont
  {Dalla~Torre}}, \bibinfo {author} {\bibfnamefont {S.~D.}\ \bibnamefont
  {Huber}},\ and\ \bibinfo {author} {\bibfnamefont {E.}~\bibnamefont
  {Altman}},\ }\bibfield  {title} {\bibinfo {title} {Nonlocal order in
  elongated dipolar gases},\ }\href
  {https://doi.org/10.1103/PhysRevB.85.125121} {\bibfield  {journal} {\bibinfo
  {journal} {Phys. Rev. B}\ }\textbf {\bibinfo {volume} {85}},\ \bibinfo
  {pages} {125121} (\bibinfo {year} {2012})}\BibitemShut {NoStop}%
\bibitem [{\citenamefont {Thorngren}\ \emph {et~al.}(2021)\citenamefont
  {Thorngren}, \citenamefont {Vishwanath},\ and\ \citenamefont
  {Verresen}}]{Thorngren2021}%
  \BibitemOpen
  \bibfield  {author} {\bibinfo {author} {\bibfnamefont {R.}~\bibnamefont
  {Thorngren}}, \bibinfo {author} {\bibfnamefont {A.}~\bibnamefont
  {Vishwanath}},\ and\ \bibinfo {author} {\bibfnamefont {R.}~\bibnamefont
  {Verresen}},\ }\bibfield  {title} {\bibinfo {title} {Intrinsically gapless
  topological phases},\ }\href {https://doi.org/10.1103/PhysRevB.104.075132}
  {\bibfield  {journal} {\bibinfo  {journal} {Phys. Rev. B}\ }\textbf {\bibinfo
  {volume} {104}},\ \bibinfo {pages} {075132} (\bibinfo {year}
  {2021})}\BibitemShut {NoStop}%
\bibitem [{\citenamefont {Fraxanet}\ \emph {et~al.}(2022)\citenamefont
  {Fraxanet}, \citenamefont {Gonz\'alez-Cuadra}, \citenamefont {Pfau},
  \citenamefont {Lewenstein}, \citenamefont {Langen},\ and\ \citenamefont
  {Barbiero}}]{Fraxanet2022}%
  \BibitemOpen
  \bibfield  {author} {\bibinfo {author} {\bibfnamefont {J.}~\bibnamefont
  {Fraxanet}}, \bibinfo {author} {\bibfnamefont {D.}~\bibnamefont
  {Gonz\'alez-Cuadra}}, \bibinfo {author} {\bibfnamefont {T.}~\bibnamefont
  {Pfau}}, \bibinfo {author} {\bibfnamefont {M.}~\bibnamefont {Lewenstein}},
  \bibinfo {author} {\bibfnamefont {T.}~\bibnamefont {Langen}},\ and\ \bibinfo
  {author} {\bibfnamefont {L.}~\bibnamefont {Barbiero}},\ }\bibfield  {title}
  {\bibinfo {title} {Topological quantum critical points in the extended
  bose-hubbard model},\ }\href {https://doi.org/10.1103/PhysRevLett.128.043402}
  {\bibfield  {journal} {\bibinfo  {journal} {Phys. Rev. Lett.}\ }\textbf
  {\bibinfo {volume} {128}},\ \bibinfo {pages} {043402} (\bibinfo {year}
  {2022})}\BibitemShut {NoStop}%
\bibitem [{SM()}]{SM}%
  \BibitemOpen
  \href@noop {} {}\bibinfo {note} {See the Supplemental Material (at the url
  provided by the publisher) for further details on the effective $t$-$J$-like
  model explaining the floating phase, and additional results concerning models
  with up to next-to-NN interactions, and models with the rung hopping weaker
  than the leg hopping.}\BibitemShut {Stop}%
\bibitem [{\citenamefont {Hauschild}\ and\ \citenamefont
  {Pollmann}(2018)}]{TeNPy2018}%
  \BibitemOpen
  \bibfield  {author} {\bibinfo {author} {\bibfnamefont {J.}~\bibnamefont
  {Hauschild}}\ and\ \bibinfo {author} {\bibfnamefont {F.}~\bibnamefont
  {Pollmann}},\ }\bibfield  {title} {\bibinfo {title} {{Efficient numerical
  simulations with Tensor Networks: Tensor Network Python (TeNPy)}},\ }\href
  {https://doi.org/10.21468/SciPostPhysLectNotes.5} {\bibfield  {journal}
  {\bibinfo  {journal} {SciPost Phys. Lect. Notes}\ ,\ \bibinfo {pages} {5}}
  (\bibinfo {year} {2018})}\BibitemShut {NoStop}%
\bibitem [{\citenamefont {Pollmann}\ \emph {et~al.}(2010)\citenamefont
  {Pollmann}, \citenamefont {Turner}, \citenamefont {Berg},\ and\ \citenamefont
  {Oshikawa}}]{Pollmann2010}%
  \BibitemOpen
  \bibfield  {author} {\bibinfo {author} {\bibfnamefont {F.}~\bibnamefont
  {Pollmann}}, \bibinfo {author} {\bibfnamefont {A.~M.}\ \bibnamefont
  {Turner}}, \bibinfo {author} {\bibfnamefont {E.}~\bibnamefont {Berg}},\ and\
  \bibinfo {author} {\bibfnamefont {M.}~\bibnamefont {Oshikawa}},\ }\bibfield
  {title} {\bibinfo {title} {Entanglement spectrum of a topological phase in
  one dimension},\ }\href {https://doi.org/10.1103/PhysRevB.81.064439}
  {\bibfield  {journal} {\bibinfo  {journal} {Phys. Rev. B}\ }\textbf {\bibinfo
  {volume} {81}},\ \bibinfo {pages} {064439} (\bibinfo {year}
  {2010})}\BibitemShut {NoStop}%
\bibitem [{\citenamefont {Ejima}\ \emph {et~al.}(2014)\citenamefont {Ejima},
  \citenamefont {Lange},\ and\ \citenamefont {Fehske}}]{Fehske2014}%
  \BibitemOpen
  \bibfield  {author} {\bibinfo {author} {\bibfnamefont {S.}~\bibnamefont
  {Ejima}}, \bibinfo {author} {\bibfnamefont {F.}~\bibnamefont {Lange}},\ and\
  \bibinfo {author} {\bibfnamefont {H.}~\bibnamefont {Fehske}},\ }\bibfield
  {title} {\bibinfo {title} {Spectral and entanglement properties of the
  bosonic haldane insulator},\ }\href
  {https://doi.org/10.1103/PhysRevLett.113.020401} {\bibfield  {journal}
  {\bibinfo  {journal} {Phys. Rev. Lett.}\ }\textbf {\bibinfo {volume} {113}},\
  \bibinfo {pages} {020401} (\bibinfo {year} {2014})}\BibitemShut {NoStop}%
\bibitem [{\citenamefont {Calabrese}\ and\ \citenamefont
  {Lefevre}(2008)}]{Calabrese2008}%
  \BibitemOpen
  \bibfield  {author} {\bibinfo {author} {\bibfnamefont {P.}~\bibnamefont
  {Calabrese}}\ and\ \bibinfo {author} {\bibfnamefont {A.}~\bibnamefont
  {Lefevre}},\ }\bibfield  {title} {\bibinfo {title} {Entanglement spectrum in
  one-dimensional systems},\ }\href
  {https://doi.org/10.1103/PhysRevA.78.032329} {\bibfield  {journal} {\bibinfo
  {journal} {Phys. Rev. A}\ }\textbf {\bibinfo {volume} {78}},\ \bibinfo
  {pages} {032329} (\bibinfo {year} {2008})}\BibitemShut {NoStop}%
\bibitem [{\citenamefont {Gammelmark}\ and\ \citenamefont
  {Zinner}(2013)}]{Gammelmark2013}%
  \BibitemOpen
  \bibfield  {author} {\bibinfo {author} {\bibfnamefont {S.}~\bibnamefont
  {Gammelmark}}\ and\ \bibinfo {author} {\bibfnamefont {N.~T.}\ \bibnamefont
  {Zinner}},\ }\bibfield  {title} {\bibinfo {title} {Dipoles on a two-leg
  ladder},\ }\href {https://doi.org/10.1103/PhysRevB.88.245135} {\bibfield
  {journal} {\bibinfo  {journal} {Phys. Rev. B}\ }\textbf {\bibinfo {volume}
  {88}},\ \bibinfo {pages} {245135} (\bibinfo {year} {2013})}\BibitemShut
  {NoStop}%
\bibitem [{\citenamefont {Huse}\ and\ \citenamefont
  {Fisher}(1982)}]{Fisher1982}%
  \BibitemOpen
  \bibfield  {author} {\bibinfo {author} {\bibfnamefont {D.~A.}\ \bibnamefont
  {Huse}}\ and\ \bibinfo {author} {\bibfnamefont {M.~E.}\ \bibnamefont
  {Fisher}},\ }\bibfield  {title} {\bibinfo {title} {Domain walls and the
  melting of commensurate surface phases},\ }\href
  {https://doi.org/10.1103/PhysRevLett.49.793} {\bibfield  {journal} {\bibinfo
  {journal} {Phys. Rev. Lett.}\ }\textbf {\bibinfo {volume} {49}},\ \bibinfo
  {pages} {793} (\bibinfo {year} {1982})}\BibitemShut {NoStop}%
\end{thebibliography}%

\cleardoublepage
\appendix

\setcounter{equation}{0}
\setcounter{figure}{0}
\setcounter{table}{0}
\makeatletter
\renewcommand{\theequation}{S\arabic{equation}}
\renewcommand{\thefigure}{S\arabic{figure}}

%
\setcounter{equation}{0}
\setcounter{figure}{0}
\setcounter{table}{0}
\makeatletter
\renewcommand{\thefigure}{S\arabic{figure}}
\section{Supplementary Information}

\section{Effective $t$-$J$-like model}

In the following, and for simplicity of the discussion we consider only up to next-to-nearest-neighbor~(NNN)
interaction, i.e. only up to the term $V_{1,1}$. We assume that the dipoles are oriented along $y$. 
We denote $t$ as the leg hopping, and $t_\perp$ as the rung hopping. Note that in the main text, we have assumed them equal for simplicity, but they do not need to be so, and this may have relevant consequences, as discussed below.

\begin{figure} [h!]
\begin{center}
\includegraphics[width=\columnwidth]{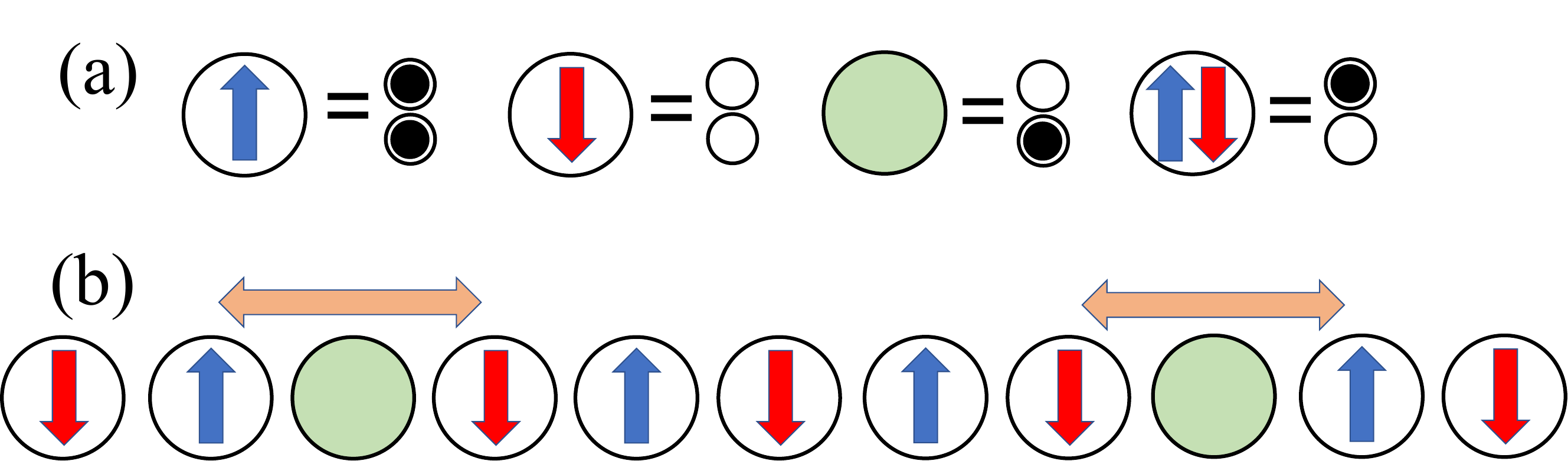}
\caption{(a) Transformation from the original ladder model to the effective $t$-$J$-like model discussed in the text. 
(b) Spins are mobile, and hence empty places can be displaced along the lattice.}
\label{fig:S1}
\end{center}
\end{figure}

The floating phase is best understood by considering an effective model in which we denote 
as particles $\uparrow$~($\downarrow$) the particles~(holes) in the upper~(lower) leg. 
Sites with neither a particle in the upper leg nor a hole in the lower one are considered as empty, see 
Fig.~\ref{fig:S1}~(a). 
Introducing the operators $\hat a_{j,\uparrow}=\hat b_{j,2}$, $\hat a_{j,\downarrow}=\hat b_{j,1}^\dag$, 
and $\hat n_{j,\uparrow}=\hat n_{j,2}$, and $\hat n_{j,\downarrow}=1-\hat n_{j,1}$, the 
Hamiltonian acquires the form:
\begin{eqnarray}
    H&=&-t\sum_j\!\sum_{\sigma=\uparrow,\downarrow} \left ( \hat{a}^\dagger_{j,\sigma}\hat{a}_{j+1,\sigma}+\mathrm{H.c.}\!\right ) 
    \nonumber \\
    &-& t_\perp \left ( \hat{a}^\dagger_{j,\uparrow} \hat{a}^\dagger_{j,\downarrow}+\mathrm{H.c.} \!\right )
     \nonumber \\
    &-& \! \frac{1}{2}\left (4V+\frac{V}{2^{3/2}} - \Delta\right )\sum_j \left ( \hat{n}_{j,\uparrow} + \hat{n}_{j,\downarrow}\right )
    \nonumber\\
    &+&
    V \sum_{j,\sigma}  \hat{n}_{j,\sigma}\hat{n}_{j+1,\sigma}
    +2V \! \sum_{j}  \hat{n}_{j,\uparrow}\hat{n}_{j,\downarrow} 
    \nonumber\\
    &+&\frac{V}{2^{5/2}} \sum_j \left ( \hat{n}_{j,\uparrow}\hat{n}_{j+1,\downarrow}
    +\hat{n}_{j,\downarrow}\hat{n}_{j+1,\uparrow} \right ).
\label{eq:H_ph}
\end{eqnarray}



We are interested in the regime where $t,t_\perp\ll V$, and $\Delta\simeq 4V$. Under these conditions, 
the term $2V \! \sum_{j}  \hat{n}_{j,\uparrow}\hat{n}_{j,\downarrow} $ introduces a strong on-site repulsion
between $\uparrow$ and $\downarrow$ particles. We may hence reduce to the hard-core regime, in which 
each rung $j$ is either empty, or occupied by an $\uparrow$ or a $\downarrow$ particle.
The model reduces to a $t$-$J$-like model of mobile spins, see Fig.~\ref{fig:S1}~(b). Note that the term 
$V \sum_{j,\sigma}  \hat{n}_{j,\sigma}\hat{n}_{j+1,\sigma}$ introduces a strong NN repulsion. As a result, 
the ground-state for $t\ll V$ is characterized by a diluted antiferromagnetic character, i.e. the positions of the spins are not fixed, but their relative orientation is Ne\'el-like. 

\begin{figure} [h!]
\begin{center}
\includegraphics[width=0.4\columnwidth]{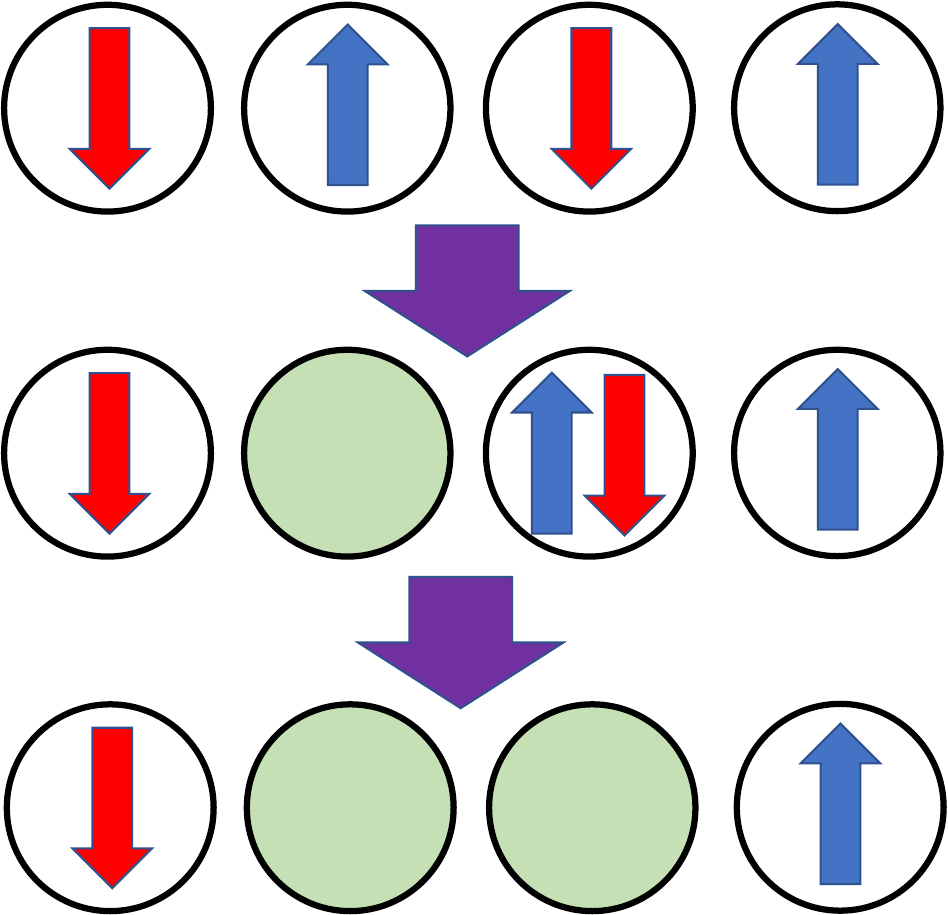}
\caption{Super-exchange process that results in the production of two neighboring empty rungs.}
\label{fig:S2}
\end{center}
\end{figure}

For $t, t_\perp \ll V$, we may generally neglect higher order terms in $t^2/V$, $t_\perp^2/V$ and $t t_\perp/V$. There is however a second-order term that may play a significant role. In the presence of 
transversal hopping $t_\perp\neq 0$, we have second-order super-exchange processes 
due to the combination of longitudinal and transversal hopping, see Fig.~\ref{fig:S2}, 
in which the number of empty rungs changes by $\pm 2$. These processes are of the order of $\frac{t t_\perp}{V}$~(note that depending on the occupation of neighboring rungs, it may be actually a bit different ranging from $\frac{t t_\perp}{V}$ to $\frac{2t t_\perp}{3V}$, but we simplify here). 



Since the spins are fixed we may focus on the overall occupation, neglecting the spin degree of freedom.
\begin{eqnarray}
    H&=&-t\sum_j \left ( \hat{b}^\dagger_{j}\hat{b}_{j+1}+{\mathrm H.c.}\!\right ) 
    \nonumber\\ &-& \! \frac{1}{2}\left (4V+\frac{V}{2^{3/2}} - \Delta\right )\sum_j n_j
    \nonumber \\
    &+& \frac{V}{2^{5/2}} \sum_j  \hat{n}_j\hat{n}_{j+1}.
\label{eq:H_ph_NNN}
\end{eqnarray}

Introducing a particle-hole transformation, $\hat c_j = \hat b_j^\dag$ and $\hat N_j = 1- \hat n_j$, 
and adding the second-order creation/destruction of empty places discussed above,  
we obtain the holon Hamiltonian of Eq. (6) of the main text.



\begin{figure*}[t!]
\begin{center}
\includegraphics[width=0.9\linewidth]{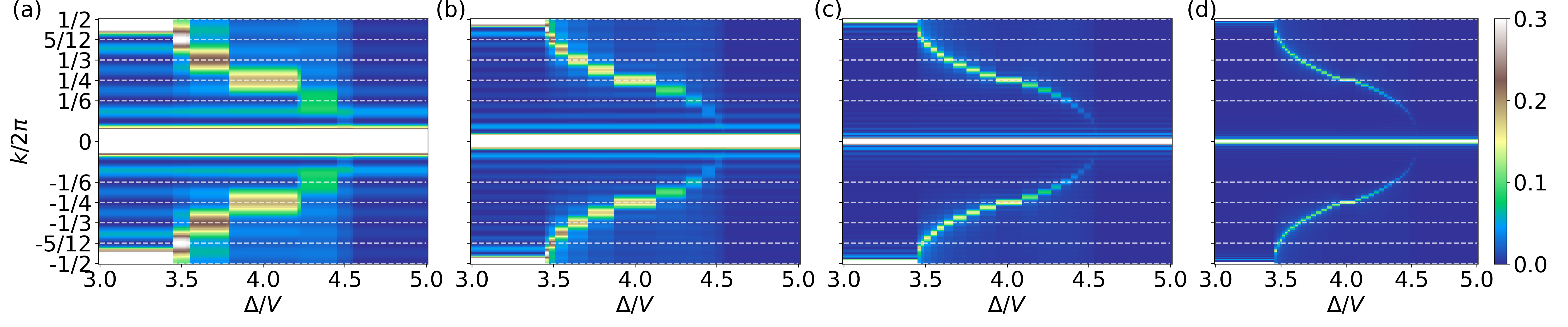}
\caption{Structure factor $S(k)$ for $t/V=0.05$ as a function of $\Delta/V$ for $L=12$ (a), 24 (b), 48 (c) and 96 (d) rungs calculated using DMRG.
}
\label{fig:S3}
\end{center}
\begin{center}
\includegraphics[width=0.9\linewidth]{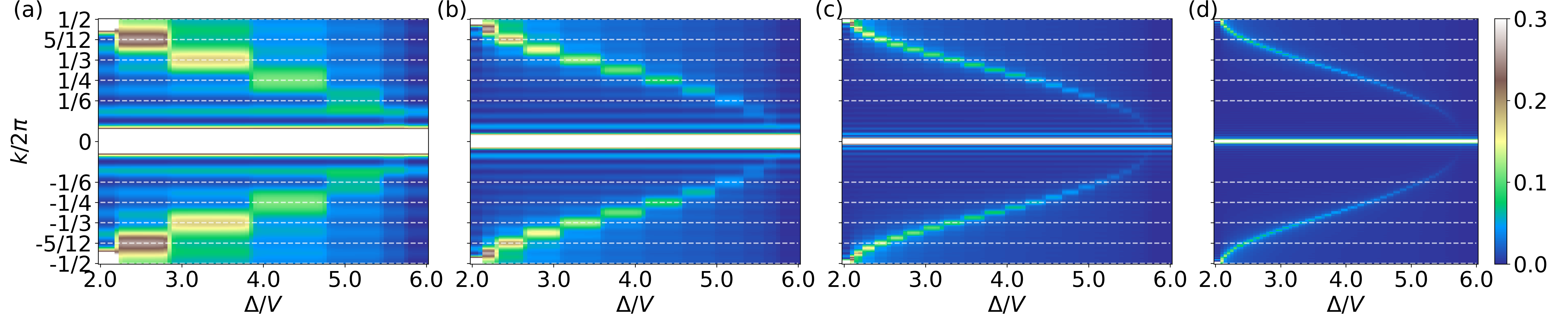}
\caption{Structure factor $S(k)$ for $t/V=0.3$ and $t_\perp/t=0.1$ as a function of $\Delta/V$ for $L=12$ (a), 24 (b), 48 (c) and 96 (d) rungs calculated using DMRG. 
}
\label{fig:S4}
\end{center}
\end{figure*}



\section{Additional results I: NNN model}

In the main text, we have shown that the floating phase, absent in the NN model, appears in the presence of the dipolar $1/r^3$ interactions. As it is clear from the effective $t$-$J$-like model discussed above, we expect that the floating phase only requires NNN interactions. We have checked that this is indeed the case. Figure~\ref{fig:S3} shows our results for the same case of Fig. 3 of the main text, but considering interactions only up to NNN, showing the clear appearance of the floating phase. We note as well the presence of the 4DW phase (plateau at $k=\pi/2$). Finally, we note that at $t=0$ (not shown) there is no 3DW phase for the NNN model.

\section{Additional results II: $t_\perp/t < 1$}

In the main text we have assumed for simplicity that the rung and leg hoppings are the same, i.e. $t_\perp=t$. This is however not necessarily the case, and it may have relevant consequences. As it becomes clear from the discussion of the $t$-$J$-like model above, 
we expect that reducing $t_\perp/t$ enhances the floating phase, since it will handicap the creation of uncorrelated holon pairs. We have checked that this is indeed the case. Figure~\ref{fig:S4} shows our results for $t/V = 0.3$ for $t_\perp/t=0.1$. Whereas for $t_\perp=t$ and $t/V=0.3$ there is no floating phase, but rather the Haldane phase, in between the 2DW and P phases, for $t_\perp/t=0.1$ a clear floating phase develops. Note that there is no plateau at $k=\pi/2$ as in Fig. 3 of the main text or in Fig.~\ref{fig:S3} because for that large value of $t/V$ there is no 4DW any more.

\end{document}